\documentclass[english,aps,pra,reprint,superscriptaddress,showpacs, longbibliography, footnoteinbib]{revtex4-2}
\usepackage[utf8]{inputenc}
\usepackage{xcolor}
\usepackage{babel}
\usepackage{float}
\usepackage{bm}
\usepackage{amsmath}
\usepackage{amssymb}
\usepackage{graphicx}
\usepackage{wasysym}
\usepackage{esint}
\usepackage[bookmarks=false,
 breaklinks=false,pdfborder={0 0 1},backref=false,colorlinks=true]
 {hyperref}
\hypersetup{pdftitle={Quasiperiodicity-induced enhancement of superconductivity in one-dimensional critical systems},
 pdfauthor={Ricardo Oliveira},
 plainpages=false,pdfpagelabels,linkcolor=blue,urlcolor=blue,citecolor=blue,pdfdisplaydoctitle=true,pdfduplex=DuplexFlipLongEdge}

\makeatletter
\usepackage{color}
\usepackage{bm}
\usepackage{indentfirst}
\usepackage{hyperref}

\makeatother

\begin{document}
\title{Quasiperiodicity-induced enhancement of superconductivity in one-dimensional
critical systems}
\author{Ricardo Oliveira}
\affiliation{Centro de Física das Universidades do Minho e do Porto, LaPMET, Departamento
de Física e Astronomia, Faculdade de Ciências, Universidade do Porto,
4169-007 Porto, Portugal}
\author{Miguel Gonçalves}
\affiliation{Princeton Center for Theoretical Science, Princeton University, Princeton,
NJ 08544}
\author{Pedro Ribeiro}
\affiliation{Centro de Física e Engenharia de Materiais Avançados, LaPMET, Instituto
Superior Técnico, Universidade de Lisboa, Av. Rovisco Pais, 1049-001
Lisboa, Portugal}
\affiliation{Beijing Computational Science Research Center, Beijing 100193, China}
\author{Eduardo V. Castro}
\affiliation{Centro de Física das Universidades do Minho e do Porto, LaPMET, Departamento
de Física e Astronomia, Faculdade de Ciências, Universidade do Porto,
4169-007 Porto, Portugal}
\affiliation{Beijing Computational Science Research Center, Beijing 100193, China}
\author{Bruno Amorim}
\affiliation{Centro de Física das Universidades do Minho e do Porto, LaPMET, Departamento
de Física e Astronomia, Faculdade de Ciências, Universidade do Porto,
4169-007 Porto, Portugal}
\begin{abstract}
We show that quasiperiodicity can enhance superconductivity in one-dimensional
narrow-band systems with $s$-wave pairing. Using a generalized Aubry-André-Harper
model featuring quasiperiodic modulations in both the on-site potential
and the hoppings, we study superconductivity across the extended,
critical, and localized phases present in the noninteracting limit.
Our results show that within the critical and localized phases, the
superconducting critical temperature exhibits an algebraic scaling
with the interaction strength, in contrast to the conventional BCS
scaling observed in periodic approximants with a similar density of
states. This direct comparison establishes that the enhancement originates
from the nature of the single-particle eigenstates, rather than from
band flattening alone. Furthermore, we find that the superfluid weight
remains finite across all phases, including the localized regime,
highlighting that superconducting phase coherence is retained throughout.
The effects of quasiperiodicity on the superfluid weight are most
pronounced in the weak-coupling regime, indicating the nontrivial
role of the localization properties of the parent eigenstates in shaping
the superconducting phases.
\end{abstract}
\maketitle

\section{Introduction}

The quantum phases of quasiperiodic matter have attracted a great
deal of interest in recent years due to cold atom experiments on Bose
glasses and many-body localization (MBL) on the one hand \citep{Roati2008,Schreiber2015,Bordia2016,Bordia2017,Lueschen2017,PhysRevLett.126.040603,PhysRevLett.122.170403},
and the remarkable phenomena observed in moiré systems, which are
generically quasiperiodic, on the other \citep{Balents2020,Uri2023}.

Quasiperiodicity is responsible for a plethora of interesting physics,
with the paradigmatic example being the dramatic change in the localization
properties of the electronic wave function with \citep{Iyer2013,Khemani2017,Lee2017,PhysRevA.92.041601,PhysRevLett.115.230401,znidaric_pnas_2018,PhysRevResearch.1.032039,PhysRevB.100.104203,PhysRevLett.128.146601}
and without \citep{Aubry1980,Sokoloff1980,Kohmoto1983,Thouless1983,Kohmoto1986,Tang1986,Sinai1987,Han1994,Chang1997,Takada2004,Liu2014,Wang2016,Zeng2016,Devakul2017,Liu2017,Yahyavi2019}
interactions. In one dimension (1D), quasiperiodic patterns can change
the nature of the wave functions from plane-wave-like extended into
exponentially localized. At the delocalized-localized critical point,
the wave functions are very exotic, exhibiting multifractal properties
\citep*{Evers2008}. The Aubry-André-Harper (AAH) model \citep{Aubry1980}
is the prototypical model where the nontrivial localization properties
mentioned above can be observed and studied. The quasiperiodic sinusoidal
modulation of the on-site potential results in a spectrum-wide quantum
criticality (SWQC), such that all the single-particle eigenstates
undergo a delocalized-localized transition at a finite value of the
potential strength. This finite critical value is determined by the
model's self-duality, resulting in a transition quite distinct from
typical Anderson localization that occurs for any finite value of
the uncorrelated disorder potential \citep{Abrahams1979,Vollhardt1980,MacKinnon1981,Kramer1993}.
Significant effort has been devoted to comprehending the physics of
single-particle localization in 1D quasiperiodic systems described
by the Aubry-André model and generalizations that contain mobility
edges and/or critical phases, where the eigenstates are multifractal
at non-fine-tuned extended regions of the phase diagram \citep{PhysRevLett.104.070601,Liu2014,PhysRevLett.114.146601,Wang2016,Liu2017,Yahyavi2019,PhysRevLett.123.025301,PhysRevLett.125.073204,anomScipost}.
In the interacting case, however, the bulk of previous research on
many-body localization has centered on highly excited states in the
middle of the many-body spectrum \citep{Iyer2013,Khemani2017,Lee2017,PhysRevA.92.041601,PhysRevLett.115.230401,znidaric_pnas_2018,PhysRevResearch.1.032039,PhysRevB.100.104203,PhysRevLett.128.146601}.

The interplay between ordered phases and quasiperiodicity has also
been extensively explored \citep{PhysRevLett.83.3908,PhysRevB.65.115114,10.21468/SciPostPhys.1.1.010,PhysRevLett.126.036803,PhysRevX.7.031061,Crowley_2022}
Notably, the study of superconductivity in quasiperiodic systems has
become a rapidly advancing field, both theoretically and experimentally.
Regarding 1D quasicrystals, recent work has concentrated on the 1D
Fibonacci chain, where the on-site potential or hopping amplitudes
alternate between two distinct values based on the Fibonacci sequence.
This research covers the enhancement of intrinsic superconductivity
\citep{Sun2024,Zhu2025,Wang2024}, the Josephson and proximity effects
\citep{Sandberg2024,Rai2019,Rai2020}, and topological superconductivity
\citep{Kobialka2024}. Additionally, intrinsic superconductivity has
been studied on the standard AAH model \citep{Fan2021,Zhang2022}.
Beyond 1D quasicrystals, significant effort has been devoted to investigating
superconductivity in 2D quasicrystals, such as the Penrose and Ammann-Beenker
tilings \citep{Sakai2017,Araujo2019,Takemori2020,Ghadimi2021,hori2024self,Liu2025,Saito2025}.
Of particular importance is the first experimental discovery of superconductivity
in quasicrystals, observed in the Al--Zn--Mg quasicrystal \citep{Kamiya2018},
and the recent identification of superconductivity in a van der Waals
layered dodecagonal quasicrystal \citep{Tokumoto2024,Terashima2024}.

In the case of moiré systems, quasiperiodicity effects are often neglected
in theoretical studies of their quantum phases. However, these effects
can be of crucial importance, as recently observed experimentally
in quasiperiodic twisted trilayer graphene \citep{Uri2023}. For electronic
densities sufficiently away from charge neutrality, a quasiperiodicity-induced
high density of weakly dispersive states, with observable transport
signatures, was detected. Remarkably, superconducting phases were
found to arise precisely in this regime and their nature is still
unknown. This work brought to the limelight the importance of quasiperiodicity
in moiré materials within experimental context, which in turn begs
for a better theoretical understanding of the interplay between quasiperiodicity
and interactions. Specifically, it is important to understand the
nature of superconducting phases in the presence of strong quasiperiodic
effects.

Motivated by the aforementioned theoretical gaps, in this work we
study the interplay between superconductivity and quasiperiodicity
in a generalized version of the AAH model, that includes an off-diagonal
quasiperiodic modulation of the hoppings \citep{Han1994,Liu2014}.
Unlike the standard AAH model, which features a single critical point
separating extended and localized phases, this model exhibits an extended
critical phase, i.e., a robust critical regime spanning a wide range
of parameters. In both these phases, the wave functions are neither
fully localized nor fully extended but instead exhibit a multifractal
structure in both real and momentum space. However, it remains unclear
whether these phases share the same physical properties. For example,
in Ref.~\citep{Gonifmmodemboxccelseccfialves2024} it was reported
that interactions are irrelevant at the critical point of an extended-to-localized
transition in a quasiperiodic systems, in stark contrast with the
demonstration in Ref.~\citep{Gonçalves2024} that in a system with
an extended critical phase interactions can give rise to a novel nontrivial
ordered phase. Additionally, the single-particle spectrum in our model
develops a narrow energy band near the Fermi level, gapped from the
rest of the spectrum (See Fig.~\ref{fig:Localization diagram and spectrums}),
making it an excellent platform for investigating how critical phases
induced by quasiperiodicity , in the presence of enhanced spectral
weight near the Fermi level, shape superconductivity. Single-particle
eigenstates with similar features have also been theoretically predicted
to arise in flat-band regimes of moiré systems \citep{Fu2020,Goncalves_2022_2DMat,PhysRevB.101.235121}.
A natural question that arises is what type of superconducting states
can emerge from the noninteracting multifractal parent state, once
interactions are included.

Recently, it was found that the 1D AAH model exhibits an enhancement
of the superconducting critical temperature at the delocalization-localization
transition. This suggests an intriguing relation between quasiperiodicity
and superconductivity \citep{Fan2021}, resembling the multifractality-induced
increase of the critical temperature for disordered systems \citep{Feigelman2007,Mukerjee2008,Feigelman2010,Burmistrov2012}.
Additionally, in systems with SWQC, a mean-field approach has shown
that the multifractality of single-particle wave functions can enhance
matrix elements, leading to a pairing amplitude that is maximized
near the delocalized-localized transition, while still retaining phase
coherence, therefore allowing for the existence of a superconductor
dome near the critical point of the system \citep{Zhang2022}. In
a 3D AAH model hosting a diffusive phase, where the wave function
delocalizes in both real and momentum space at intermediate quasiperiodic
potential, $p$-wave superconductivity was shown to be suppressed
while the conventional $s$-wave instability survives \citep{DevakulPRB2024}.

In this work, we employ a self-consistent Bogoliubov-de Gennes (BdG)
formalism \citep{DeGennes1999} to study spin-singlet superconductivity
in the aforementioned generalized AAH model. A key aspect of our approach
is the direct comparison between commensurate (periodic) and incommensurate
(quasiperiodic) structures. This is achieved by constructing periodic
systems, by repetition of smaller unit cells, with similar total sizes
and modulation period close to those of the reference quasiperiodic
systems. Such a strategy allows us to compare various observables
in the periodic and quasiperiodic limits, including the critical temperature,
the superconducting order parameter, and the superfluid weight at
zero temperature. Our findings reveal that for periodic systems, the
critical temperature has the usual Bardeen-Cooper-Schrieffer (BCS)
scaling with the interaction strength, while in the quasiperiodic
case it scales algebraically, displaying a significant enhancement
at small interaction strength. Importantly, by comparing periodic
and quasiperiodic systems with similar density of states, we ruled
out alternative explanations, such as enhancements driven solely by
an increased density of states at the Fermi level. Finally, we highlight
our analysis of the scaling of the superfluid weight, which measures
the phase coherence of the superconducting phases in our systems,
with interaction strength across the extended, localized and critical
phases of our model. We reveal that signatures of quasiperiodicity
are only evident in the weak-coupling limit. Furthermore, by performing
finite-size scaling analysis, we show that the superfluid weight saturates
to a finite value in the quasiperiodic limit, even in the critical
and localized phases of our generalized AAH model. This indicates
that all the phases exhibit superconducting phase coherence for a
finite interaction strength.

The remaining paper is organized as follows. The model, methods and
quantities studied in this work are described in Sec.~\ref{sec:Model-and-Methods}.
The obtained results are described and analyzed in Sec.~\ref{sec:Results}
and conclusion are drawn in Sec.~\ref{sec:Discussion}. Details on
some derivations and numerical details are provided as Appendix.

\section{Model and Methods\label{sec:Model-and-Methods}}

\subsection{Model and BdG Hamiltonian}

We will study superconductivity on the generalized 1D AAH model, which
is described by the following Hamiltonian \citep{Han1994,Liu2014}
\begin{align}
\hat{H}_{0}= & -\sum_{\sigma,m}\left[t+V_{2}\cos\left(2\pi Qm+\pi Q\right)\right]c^{\dagger}_{m+1,\sigma}c_{m,\sigma}+\text{H.c.}\nonumber \\
 & +\sum_{\sigma,m}\left[V_{1}\cos\left(2\pi Qm\right)-\mu\right]c^{\dagger}_{m,\sigma}c_{m,\sigma}.\label{eq:Generalized AAH model}
\end{align}
where the first term contains nearest-neighbor hoppings, $t$, with
an additional sinusoidal modulation of amplitude $V_{2}$ and wavenumber
$2\pi Q$, the second term contains a modulated on-site potential
of amplitude $V_{1}$ and the same wavenumber $2\pi Q$, and $\mu$
is the chemical potential. By setting $V_{2}=0$, one recovers the
usual AAH model. The model can also be seen as the reduction of a
2D Quantum Hall system with isotropic next-nearest-neighbor hoppings
to a 1D chain. From this point on, we will take $t=1$, i.e., all
energies will be measured in units of $t$.

\noindent If $Q$ is a rational number, $Q=p/q$ with $p$ and $q$
integer coprime numbers, then the model defined by Eq.~\eqref{eq:Generalized AAH model}
is periodic, with a unit cell with $q$ sites. On the other hand,
if $Q$ is an irrational number then the model is quasiperiodic, lacking
any discrete translational symmetry. In this work, we take $Q$ to
be the inverse of the golden ratio, $\varphi^{-1}=\left(\sqrt{5}-1\right)/2$,
for our studies of the quasiperiodic case. In the numerical simulations,
we will consider finite systems, and in order to minimize finite size
effects, we will impose periodic boundary conditions. Quasiperiodic
systems are then simulated by considering rational approximants to
the irrational number $Q$. The rational approximants to $\varphi^{-1}$
are know to be given by ratios of Fibonacci numbers $Q_{n}=F_{n-1}/F_{n}$,
where the Fibonacci sequence $\left\{ F_{n}\right\} $ is defined
recursively via the relation $F_{n}=F_{n-1}+F_{n-2}$ with $F_{0}=F_{1}=1$.
The irrational number $\varphi^{-1}$ is recovered by taking the limit
$\varphi^{-1}=\lim_{n\rightarrow\infty}Q_{n}$. For a fixed $Q_{n}$
the quasiperiodic system is simulated by considering a system with
$N=F_{n}$ sites. To compare periodic and quasiperiodic systems, we
consider systems with similar sizes but with different choices of
rational approximants. Specifically, if we consider a quasiperiodic
system with $Q_{n}$ and system size $N=F_{n}$, we examine systems
with lower order rational approximants, $i<n$, where the modulation
frequency is given by $Q_{i}=F_{i-1}/F_{i}$. These systems have a
unit cell of size $F_{i}$ which is repeated $N_{\text{u.c.}}$ times,
such that the size of the periodic systems, $N_{i}=N_{\text{u.c.}}F_{i}$,
approximates the size of the quasiperiodic one $N=F_{n}$. This procedure
ensures a fair comparison between periodic and quasiperiodic systems
of similar size and modulation wavenumber. For a given target $N=F_{n}$,
we interpolate between the quasiperiodic and the periodic limits by
considering systems with lower order approximants, $Q_{i}$, and increasing
the number of repeated unit cells, $N_{\text{u.c.}}$, such that $F_{i}N_{\text{u.c.}}\simeq N$.

\begin{figure}
\begin{centering}
\includegraphics[scale=0.25]{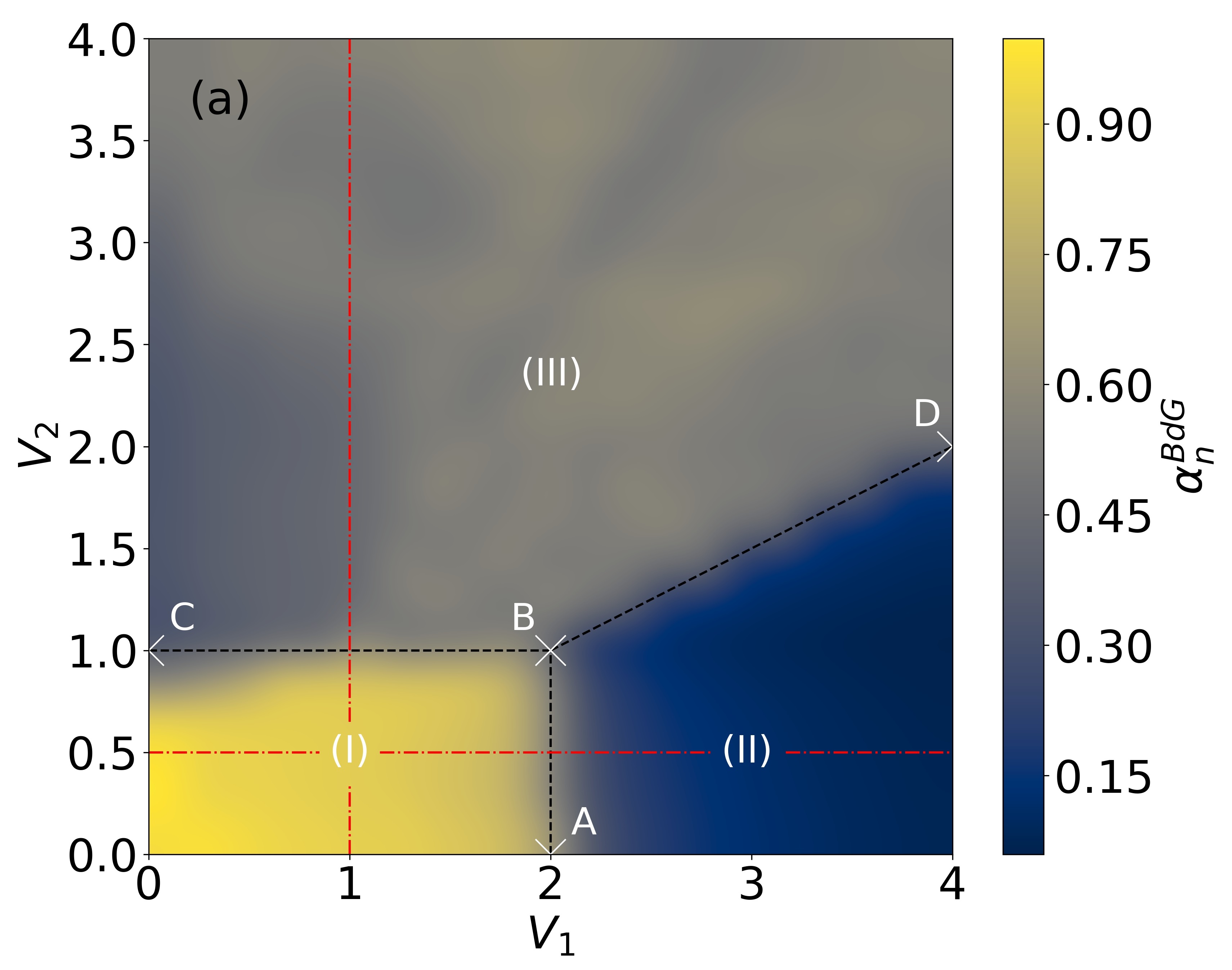}
\par\end{centering}
\begin{centering}
\includegraphics[scale=0.2]{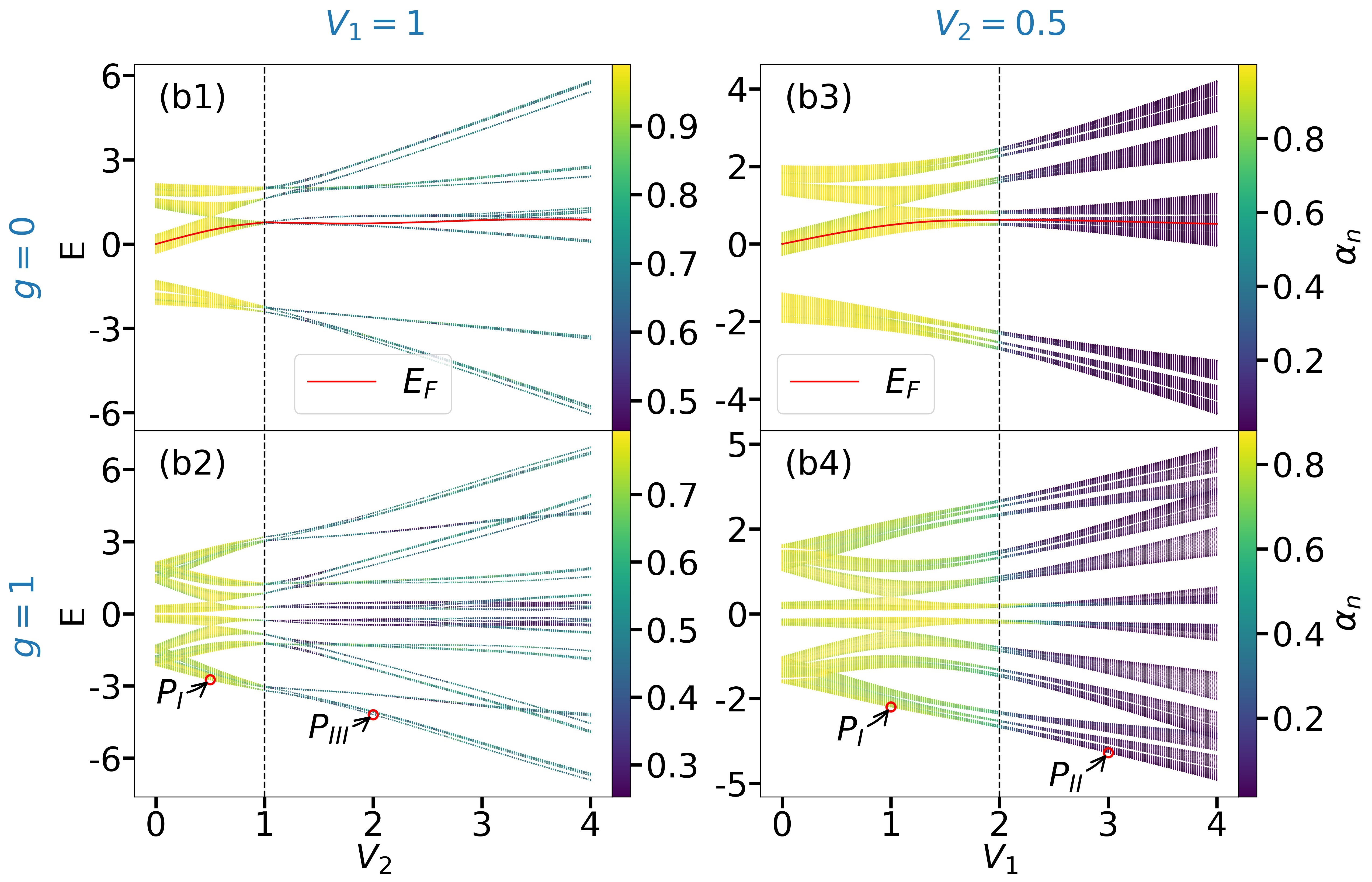}
\par\end{centering}
\caption{(a) The MIPR exponent $\alpha^{\text{BdG}}_{n}$, defined in Eq.~\eqref{eq:MIPR exponent},
for the BdG eigenstates in the $\left(V_{1},V_{2}\right)$ parameter
space at zero temperature, with $g=1$ for a system of size $N=610$.
$\alpha^{\textrm{BdG}}_{n}$ converges to $1$ for extended states,
$0$ for localized states and to noninteger in-between values for
critical states as $N\rightarrow\infty$. The free system transition
lines $\overline{AB}$, $\overline{BC}$, and $\overline{BD}$ are
plotted as dashed black lines, with region~I the extend phase, region~II
the localized phase, and region~III the critical phase. (b) Spectrum
along representative (red-dashed) lines shown in panel~(a): panels~(b1)
and~(b3) are for $\hat{H}_{0}$ (free system), respectively at fixed
$V_{1}=1$ and at fixed $V_{2}=0.5$, for size $N=2584$; panels~(b2)
and~(b4) are for $\hat{H}_{\text{BdG}}$ with $g=1$, for size $N=610$.
We represent, in color, the $n$-th order approximation to the IPR
exponent $\alpha$, given by $\alpha_{n}\equiv-\log\left[\text{IPR}\right]/\log\left[F_{n}\right]$.\label{fig:Localization diagram and spectrums}}
\end{figure}

The localization phase diagram of the model in Eq.~\ref{eq:Generalized AAH model}
in the plane $\left(V_{1},V_{2}\right)$ for the incommensurate system
was first studied in Ref.~\citep{Han1994}. For any energy, the eigenstates
are extended for $\left\{ V_{1}<2t\:\land\:V_{2}<t\right\} $, localized
for $\left\{ V_{1}>2\max\left(t,V_{2}\right)\right\} $, and critical
for $\left\{ V_{2}>\frac{1}{2}\max\left(2t,V_{1}\right)\right\} $.
These regions are delimited by dashed black lines in the localization
phase diagram of Fig.~\ref{fig:Localization diagram and spectrums}(a),
and labeled region~I, II, and III, respectively for the extended,
localized, and critical phase.

In the calculations that follow, we adjust the chemical potential
$\mu$ for each point $\left(V_{1},V_{2}\right)$ in the phase diagram
in order to ensure that the noninteracting system is always at half-filling.
This is particularly important for points in the diagram whose spectrum
is not symmetric and other choices of chemical potential could result
in the suppression of superconductivity simply due to gaps in the
noninteracting spectrum.

To study the effects of the incommensurate nature of the generalized
AAH model on superconductivity, we include a simple interaction term
in our Hamiltonian. Specifically, the total Hamiltonian will include
in addition to $\hat{H}_{0}$ of Eq.~\eqref{eq:Generalized AAH model}
an attractive on-site Hubbard term 
\begin{equation}
\hat{H}=\hat{H}_{0}-g\sum_{m}c^{\dagger}_{m\uparrow}c^{\dagger}_{m\downarrow}c_{m\downarrow}c_{m\uparrow},\label{eq:Full Hamiltonian}
\end{equation}
which is defined by a single interaction parameter $g$. Within a
mean-field approximation, we can define the superconducting on-site
pairing as $\Delta_{m}\equiv-g\left\langle c_{m\downarrow}c_{m\uparrow}\right\rangle $,
and if the noninteracting systems is spin degenerate, we obtain the
BdG Hamiltonian as 
\begin{equation}
\hat{H}_{\text{BdG}}=\frac{1}{2}\Psi^{\dagger}\left[\begin{array}{cccc}
\boldsymbol{h} & -\boldsymbol{\Delta} & 0 & 0\\
-\boldsymbol{\Delta}^{\dagger} & -\boldsymbol{h}^{*} & 0 & 0\\
0 & 0 & \boldsymbol{h} & \boldsymbol{\Delta}\\
0 & 0 & \boldsymbol{\Delta}^{\dagger} & -\boldsymbol{h}^{*}
\end{array}\right]\Psi,\label{eq:BdG Hamiltonian}
\end{equation}

\noindent where $\Psi\equiv\begin{pmatrix}\boldsymbol{c}_{\downarrow} & \boldsymbol{c}^{\dagger}_{\uparrow} & \boldsymbol{c}_{\uparrow} & \boldsymbol{c}^{\dagger}_{\downarrow}\end{pmatrix}^{T},$
such that $\boldsymbol{c}_{\sigma}$ $\left(\boldsymbol{c}^{\dagger}_{\sigma}\right)$
is a column vector of annihilation (creation) operators with $N$
entries, for a lattice with $N$ total sites, and we defined the matrices
$\left(\boldsymbol{h}\right)_{m,n}=H_{0;m,n}$ and $\left(\boldsymbol{\Delta}\right)_{m,n}=\Delta_{n}\delta_{m,n}$.
The BdG canonical transformation amounts to introducing an unitary
transformation that diagonalizes the first block in $\boldsymbol{H}_{\text{BdG}}$
\begin{equation}
\left[\begin{array}{cc}
\boldsymbol{h} & -\boldsymbol{\Delta}\\
-\boldsymbol{\Delta}^{\dagger} & -\boldsymbol{h}^{*}
\end{array}\right]=\left[\begin{array}{cc}
\boldsymbol{U} & -\boldsymbol{V}^{*}\\
\boldsymbol{V} & \boldsymbol{U}^{*}
\end{array}\right]\left[\begin{array}{cc}
\boldsymbol{E} & \bm{0}\\
\bm{0} & -\bm{E}
\end{array}\right]\left[\begin{array}{cc}
\boldsymbol{U}^{\dagger} & \boldsymbol{V}^{\dagger}\\
-\boldsymbol{V}^{T} & \bm{U}^{T}
\end{array}\right],\label{eq:BdG_diagonalization}
\end{equation}
where $\boldsymbol{E}$ is a $N\times N$ diagonal matrix, $E_{\mu\nu}=\delta_{\mu\nu}E_{\mu}$,
where $E_{\mu}>0$ are the positive eigenvalues. Considering the two
blocks, the diagonalized Hamiltonian takes the form
\begin{equation}
\hat{H}_{\text{BdG}}=\sum^{N-1}_{\mu=0}E_{\mu}\left(\gamma^{\dagger}_{\mu,1}\gamma_{\mu,1}+\gamma^{\dagger}_{\mu,2}\gamma_{\mu,2}\right),\label{eq:Diagonalized_BdG_Hamiltionian}
\end{equation}
where
\begin{equation}
\left[\begin{array}{c}
\boldsymbol{\gamma}_{1}\\
\boldsymbol{\gamma}^{\dagger}_{2}
\end{array}\right]=\left[\begin{array}{cc}
\boldsymbol{U}^{\dagger} & \boldsymbol{V}^{\dagger}\\
-\boldsymbol{V}^{T} & \bm{U}^{T}
\end{array}\right]\left[\begin{array}{c}
\bm{c}_{\downarrow}\\
\boldsymbol{c}^{\dagger}_{\uparrow}
\end{array}\right],\label{eq:BdG canonical transformation}
\end{equation}
with $\boldsymbol{\gamma}_{\tau}$ $\left(\boldsymbol{\gamma}^{\dagger}_{\tau}\right)$
a column vector of annihilation (creation) operators with $N$ entries.
The superconducting pairing parameters are then determined self-consistently
by using 
\begin{equation}
\Delta_{m}=-g\sum^{N-1}_{\mu=0}U_{m,\mu}V^{*}_{m,\mu}\tanh\left(\frac{E_{\mu}}{2k_{B}T}\right),\label{eq:Self-consistent equation for gap}
\end{equation}
where $U_{m,\mu}=\left[\bm{U}\right]_{m,\mu}$ and $V_{m,\mu}=\left[\bm{V}\right]_{m,\mu}$
are the entries of the matrices defining the unitary transformation
in Eq.~\eqref{eq:BdG_diagonalization}.

Our mean-field approach involves iteratively solving Eqs.~\eqref{eq:BdG_diagonalization}
and \eqref{eq:Self-consistent equation for gap}, until convergence
is attained. We use the relative error between two consecutive iterations
of the superconducting pairing vector $\boldsymbol{\Delta}$ as our
convergence criterion. The relative error is defined as 
\begin{equation}
\epsilon=\frac{\left\Vert \boldsymbol{\Delta}^{i+1}-\boldsymbol{\Delta}^{i}\right\Vert }{\left\Vert \boldsymbol{\Delta}^{i+1}\right\Vert },\label{eq:Relative error}
\end{equation}
where $\left\Vert \bm{\Delta}^{i}\right\Vert =\sqrt{\sum_{m}\left|\Delta_{m}\right|^{2}}$
is the norm of the vector $\bm{\Delta}$ in the $i$-th iteration.
We chose to terminate our algorithm when $\epsilon<10^{-4}$, which
can take up to $50$ iterations to reach convergence in the localized
phase. The mean-field approximation has been widely used to study
superconducting phases in 1D quasiperiodic systems in the literature
\citep{Fan2021,Zhang2022,Wang2024,Sun2024,Rai2019,Rai2020}. We therefore
expect it to provide qualitatively accurate results, particularly
for comparing the quasiperiodic and periodic limits of the superconducting
properties of our model, which is the primary focus of this work.

Although the BdG method is an excellent approach for analyzing superconducting
properties of 1D systems for temperatures below the superconducting
critical temperature, it is not very efficient for computing the critical
temperature itself. This is due to the large number of iterations
required to obtain convergence near the temperature-driven transition.
Instead, within the same mean-field approximation, a more appropriate
method for this task is the linearized gap equation. Following the
approach in Ref.~\citep{Fan2021}, the critical temperature can be
computed from the linearized gap equation in the Anderson approximation
\footnote{The Anderson approximation can be seen as an extension of BCS pairing
to inhomogeneous systems. In this approximation, the full Hamiltonian,
Eq.~\eqref{eq:Full Hamiltonian}, is written in the basis that diagonalizes
the single-particle Hamiltonian, and it is assumed that superconducting
pairing only occurs for time-reversed states.}, valid in the weak- coupling limit, which can be written as 
\begin{equation}
\Delta_{\lambda}=g\sum_{\lambda^{\prime}}M_{\lambda,\lambda^{\prime}}\frac{\tanh\left[\frac{E_{\lambda^{\prime}}}{2k_{B}T_{c}}\right]}{2E_{\lambda^{\prime}}}\Delta_{\lambda^{\prime}},\label{eq:Linear gap equation Anderson approximation}
\end{equation}
where $M_{\lambda,\lambda^{\prime}}\equiv\sum_{i}\left|\phi^{\lambda}_{i}\right|^{2}\left|\phi^{\lambda^{\prime}}_{i}\right|^{2}$
is computed from the eigenstates $\phi^{\lambda}_{i}$ with eigenvalue
$E_{\lambda}$ of $\boldsymbol{h}$. This is an eigenvalue problem
for the superconducting pairing $\left\{ \Delta_{\lambda}\right\} $
in the eigenbasis of $\boldsymbol{h}$, and $T_{c}$ is the temperature
for which the highest eigenvalue of the gap equation matrix becomes
unity.

Since the BdG algorithm is computationally more demanding than the
linear gap equation, we limit the BdG results to total system sizes
of $N=610$, while the $T_{c}$ results are limited to $N=2584$.
We have confirmed that these sizes are sufficient to converge all
observables, enabling a fair comparison between quasiperiodic and
periodic systems.

\subsection{Inverse participation ratio analysis of localization properties}

A useful quantity to study the localization properties of the eigenstates
of a quasiperiodic system is the inverse participation ratio (IPR),
which quantifies the fraction of the total probability density that
is concentrated on a subset of lattice sites. For the free system,
the IPR for a normalized eigenstate $\phi^{\lambda}$ corresponding
to the $E_{\lambda}$ eigenvalue is defined as 
\begin{equation}
\text{IPR}\left(E_{\lambda}\right)=\sum^{N}_{m=0}\left|\phi^{\lambda}_{m}\right|^{4}.\label{eq:Free system IPR}
\end{equation}
It is a well known result that for extended eigenstates, the IPR scales
as $N^{-1}$ in the thermodynamic limit, while for the localized states
it tends to a finite value. However, for critical multifractal states,
the IPR scales as $N^{-\alpha}$ with $\alpha\in\left]0,1\right[$
depending on the critical properties of the eigenstates.

A suitable quantity to study the localization phase diagram of the
BdG Hamiltonian is the mean inverse participation ratio (MIPR) of
the BdG wave functions, 
\begin{equation}
\text{MIPR}=\frac{1}{2N}\sum^{2N-1}_{\lambda=0}\sum^{N-1}_{m=0}\left[\left|U_{m,\lambda}\right|^{4}+\left|V_{m,\lambda}\right|^{4}\right],\label{eq:MIPR BdG}
\end{equation}
where we have averaged over BdG wave functions with both positive
and negative eigenvalues. The slope of the MIPR varies rapidly near
the transition lines between the extended, critical, or localized
phases. Therefore, we can expect a jump in this quantity in the thermodynamic
limit, which will allow us to compute the localization phase diagram
for this system. In analogy to the IPR, we will estimate a multifractal
exponent for the BdG wave functions as
\begin{equation}
\alpha^{\text{BdG}}_{n}=-\frac{\log\left(\text{MIPR}\right)}{\log\left(F_{n}\right)},\label{eq:MIPR exponent}
\end{equation}
for a system with size $N=F_{n}$.

On the other hand, if we want to investigate the effects of localization
on the superconducting pairing, we can define an IPR for the $\Delta_{m}$,
similar to what is done for the wave functions, as
\begin{equation}
\text{IPR}_{\Delta}=\frac{\sum^{N-1}_{m=0}\left|\Delta_{m}\right|^{4}}{\left[\sum^{N-1}_{m=0}\left|\Delta_{m}\right|^{2}\right]^{2}}.\label{eq:IPR_delta}
\end{equation}
This is an energy-independent quantity, which characterizes the localization
of $\Delta_{m}$ as a distribution on the lattice. A similar localization
analysis of the inverse participation ratio can be done for local
observables in total analogy to what is typically done for eigenstates,
as shown in Ref.~\citep{JANSSEN1994}. For example, the standard
IPR has been adapted to the study of the localization properties of
charge fluctuations \citep{Gonçalves2024}.

\section{Results\label{sec:Results}}

\subsection{Localization diagram\label{subsec:Localization-diagram}}

We applied the BdG method to the generalized AAH model with an on-site
Hubbard term, at zero temperature, and for moderate values of the
interaction parameter, i.e., of the order of the uniform hopping parameter
$t$. We found that it shares a similar localization phase diagram
with the parent model, as shown in Fig.~\ref{fig:Localization diagram and spectrums}(a).
The MIPR exponent for the BdG wave functions allows us to define three
phases, as in the free model: extended, critical, and localized. The
exponent value is almost uniform inside each region and varies sharply
close to the transition lines, juxtaposing on the critical lines of
the free model. The mean-field algorithm converges smoothly for all
points in the diagram, starting from a homogeneous initial proposal
of the superconducting pairing.

In Fig.~\ref{fig:Localization diagram and spectrums}(b) we plot
the spectrum along representative (red-dashed) lines shown in Fig.~\ref{fig:Localization diagram and spectrums}(a).
The free system ($g=0$) exhibits SWQC, with all eigenstates undergoing
a transition from extended to critical at $V_{2}=V^{c}_{2}=1$ for
$V_{1}=1$ {[}Fig.~\ref{fig:Localization diagram and spectrums}(b1){]},
and from extended to localized states at $V_{1}=V^{c}_{1}=2$ for
$V_{2}=0.5$ {[}Fig.~\ref{fig:Localization diagram and spectrums}(b3){]}.
However, this is only roughly true for the BdG system ($g=1$), and
we can see in Figs.~\ref{fig:Localization diagram and spectrums}(b2)
and \ref{fig:Localization diagram and spectrums}(b4) that the spectrum
exhibits some mobility edge structure. In particular, we can see in
Fig.~\ref{fig:Localization diagram and spectrums}(b4) that the states
close to the BdG gap transition from extended to localized for a critical
value $V^{c}_{1}\gtrsim2$. Additionally, in the transition from extended
to critical shown in Fig.~\ref{fig:Localization diagram and spectrums}(b2),
there is also a transition to localized states for states whose energy
lies closest to the superconducting gap.

We now turn to the spatial distribution of the superconducting pairing
with the aim to understand whether it follows the localization of
the BdG wave functions. We found that, although $\text{IPR}_{\Delta}$
is larger in the critical phase and even more so in the localized
phase, the superconducting pairing distribution never actually acquires
a critical or localized structure. We can verify this by looking at
some relevant points in the $\left(V_{1},V_{2}\right)$ diagram. These
points are representative of the general behavior within their respective
phases. We verified this by testing additional points within each
phase, obtaining qualitatively similar behaviors. This consistency
applies to both the localization properties of the BdG wave functions
(Sec.~\ref{subsec:Localization-diagram}) and the scaling of the
critical temperature (Sec.~\ref{subsec:Critical-temperature and zero-temperature pairing}).

\begin{figure}
\centering{}\includegraphics[scale=0.22]{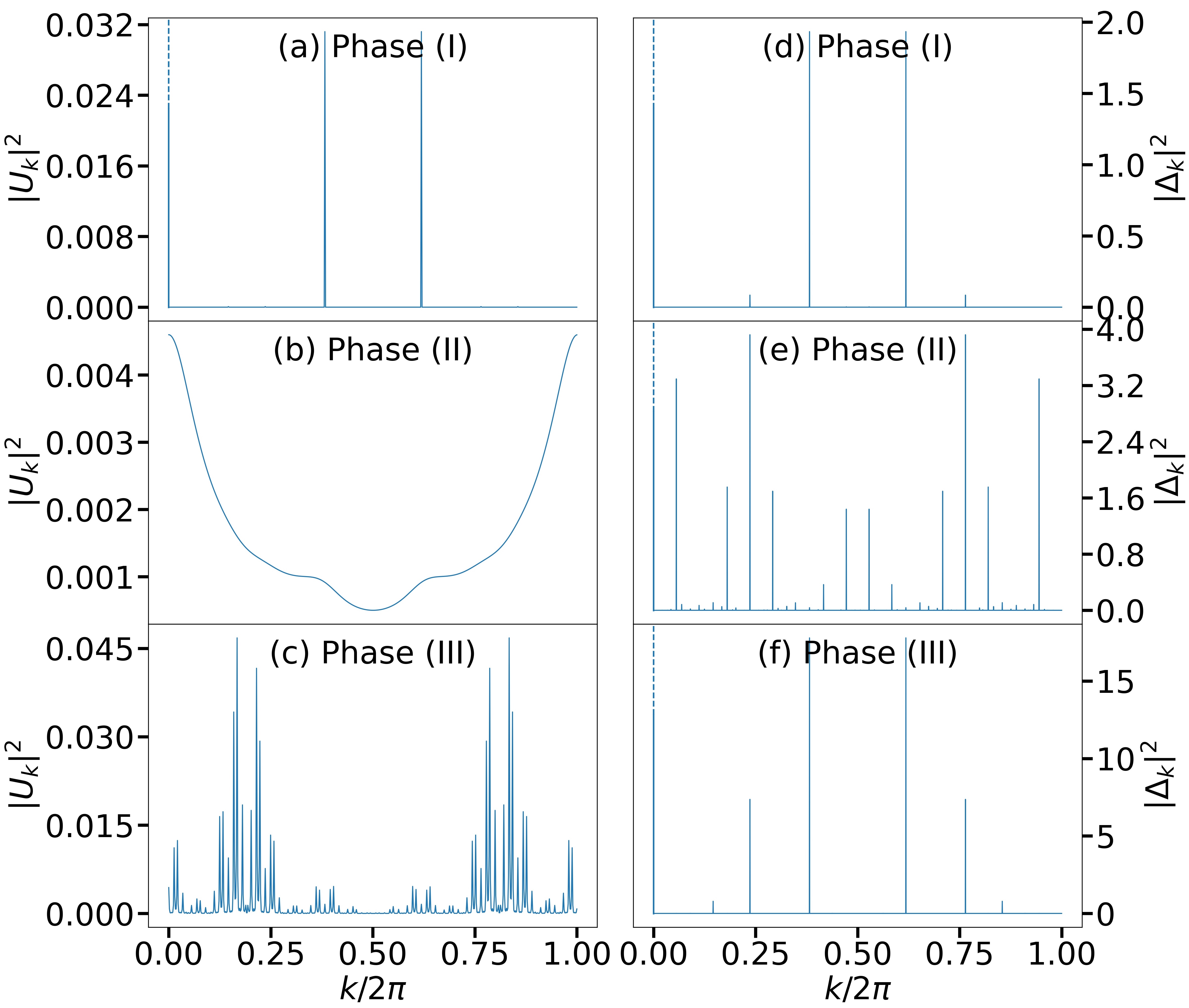}\caption{Momentum-space distribution of the BdG wave functions and of the superconducting
pairing, for a system of size $N=F_{n}=2584$ and interaction parameter
$g=1$. In (a-c) we plot the Fourier transform $\left|U_{k}\right|^{2}$
of the $U$-components of the BdG wave function at the lower edge
of the spectrum ($E=E_{\text{min}}$). The chosen points are represented
in Figs.~\ref{fig:Localization diagram and spectrums}(b2) and (b4)
as $P_{\text{I}}$, $P_{\text{II}}$, and $P_{\text{III}}$, corresponding
to the points $\left(V_{1},V_{2}\right)=\left(1,0.5\right)$ (a),
$\left(V_{1},V_{2}\right)=\left(3,0.5\right)$ (b) and $\left(V_{1},V_{2}\right)=\left(1,2\right)$
(c). In (d-f) we plot the Fourier transform $\left|\Delta_{k}\right|^{2}$
of the superconducting pairing vector for the same points in diagram
$\left(V_{1},V_{2}\right)$.\label{fig:Momentum-space-distributions}}
\end{figure}

In Figs.~\ref{fig:Momentum-space-distributions}(a)-\ref{fig:Momentum-space-distributions}(c)
we show the normalized Fourier transform of some BdG wave functions,
for the points which are marked in Fig.~\ref{fig:Localization diagram and spectrums}
as $P_{\text{I}}$, $P_{\text{II}}$, and $P_{\text{III}}$, and in
Fig.~\ref{fig:Momentum-space-distributions}(d)- the Fourier transform
of the superconducting pairing, at the same points in the diagram.
In phase~I, the behavior is the same, with both the BdG wave functions
and superconducting pairing showing the same ballistic structure,
evident in the exponentially localized peaks centered at certain crystal
momentum values shown in Figs.~\ref{fig:Momentum-space-distributions}(a)
and~\ref{fig:Momentum-space-distributions}(d). However, it is in
the critical and localized phases that we see the biggest differences
between these two scalar distributions. As can be seen in Fig.~\ref{fig:Momentum-space-distributions}(b),
in phase~II the BdG wave functions delocalize in the first Brillouin
zone, corresponding to a localization in real space. In phase~III,
the BdG wave functions undergo a partial delocalization around the
ballistic peaks, as shown in Fig.~\ref{fig:Momentum-space-distributions}(c),
corresponding to a critical distribution in real space. On the other
hand, the distribution of the superconducting pairing in momentum
space remains ballistic in these two phases, as can be inferred from
Figs.~\ref{fig:Momentum-space-distributions}(e) and~\ref{fig:Momentum-space-distributions}(f).
The larger value of $\text{IPR}_{\Delta}$ is explained only by the
appearance of more ballistic peaks for an extra small subset of momentum
values.

\begin{figure}
\centering{}\includegraphics[scale=0.29]{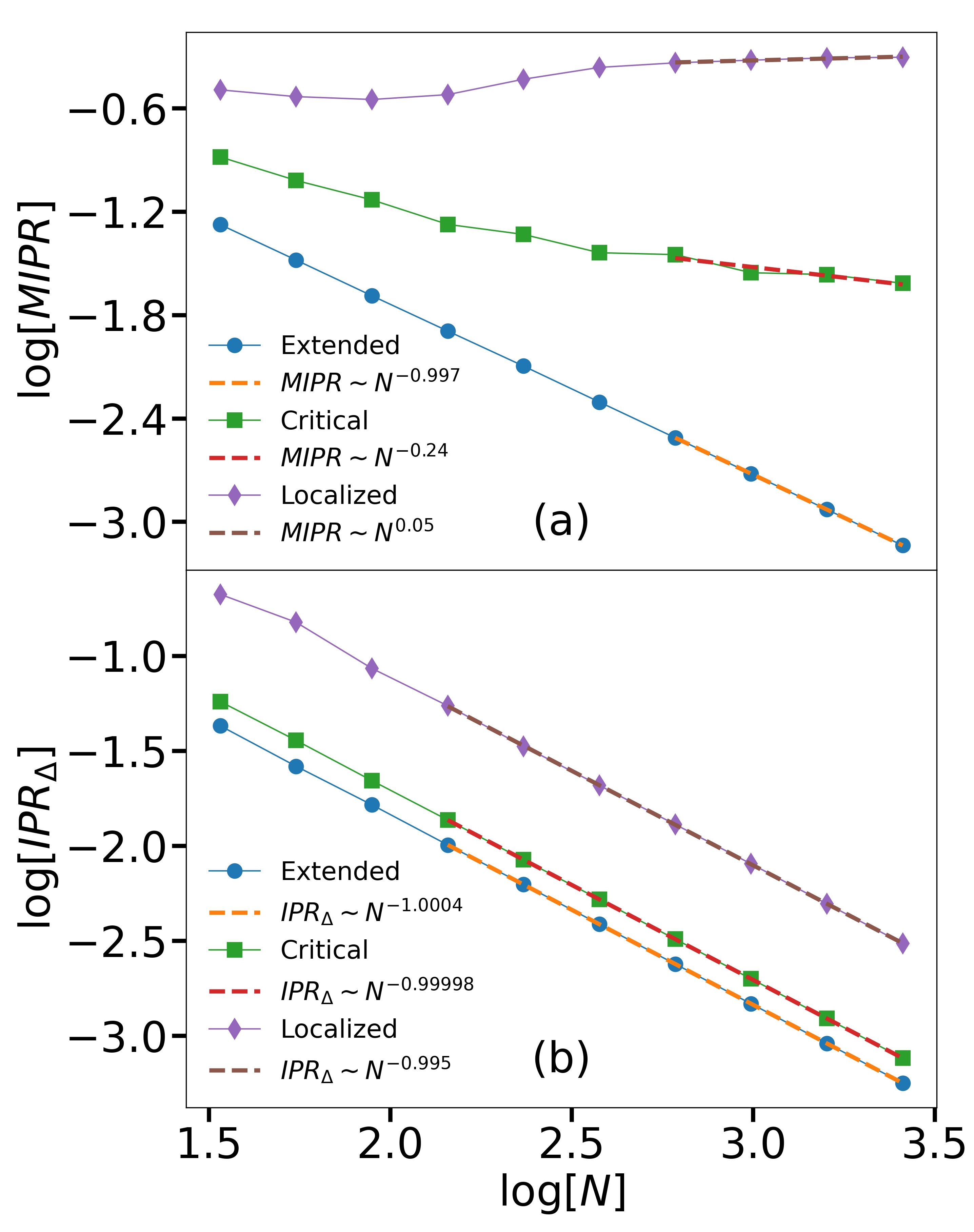}\caption{(a) Finite-size analysis of $\text{MIPR}$ for the BdG wave functions,
with $g=1$. The log-log plot of $\text{MIPR}$ as a function of the
system size $N$ for three relevant points $\left(V_{1},V_{2}\right)=\left(1,0.5\right)$
(extended phase I), $\left(V_{1},V_{2}\right)=\left(3,0.5\right)$
(localized phase II) and $\left(V_{1},V_{2}\right)=\left(1,2\right)$
(critical phase III). We perform a linear fit to $\text{IPR}\sim N^{-\alpha}$
for the larger system sizes, obtaining $\alpha^{\text{BdG}}_{I}=0.997\pm0.001$,
$\alpha^{\text{BdG}}_{II}=0.05\pm0.02$ and $\alpha^{\text{BdG}}_{III}=0.24\pm0.05$.
(b) The same analysis is done for $\text{IPR}_{\Delta}$. Even though
$\text{IPR}^{\text{I}}_{\Delta}<\text{IPR}^{\text{III}}_{\Delta}<\text{IPR}^{\text{II}}_{\Delta}$,
as expected, we found that this quantity has an extended-like scaling
in all three phases.\label{fig:Finite-size-analysis}}
\end{figure}

To complete our analysis of the localization phase diagram for our
model, we performed a finite-size analysis of both the $\text{MIPR}$
of the BdG wave functions and the $\text{IPR}_{\Delta}$ for relevant
points in Fig.~\ref{fig:Finite-size-analysis}. As mentioned in Sec.~\ref{sec:Model-and-Methods},
we are able to do a linear fit to $\text{IPR}\sim N^{-\alpha}$ in
a log-log plot of the $\text{IPR}$ against the size of the system
in the large $N$ limit, allowing us to extrapolate and estimate the
scaling exponent $\alpha$. In Fig.~\ref{fig:Finite-size-analysis}(a),
we see that we enter this algebraic regime for a system of size $N=610$,
indicating that our localization diagram in Fig.~\ref{fig:Localization diagram and spectrums}
is sufficiently converged to draw our previous conclusions. The algebraic
scaling allows us to label phases I, II, and III as extended, localized
and critical, respectively. This confirms our result that the BdG
wave functions are determined by incommensurate modulations in the
same way as the eigenstates of the free system.

In Fig.~\ref{fig:Finite-size-analysis}(b) we plot $\text{IPR}_{\Delta}$
as a function of system size in phases I, II, and III. It is noteworthy
that $\text{IPR}^{\text{I}}_{\Delta}<\text{IPR}^{\text{III}}_{\Delta}<\text{IPR}^{\text{II}}_{\Delta}$
for all values of $N$. However, $\text{IPR}_{\Delta}$ scales with
$N$ in the same way as one would expect for a ballistic extended
state. We can therefore conclude that the superconducting pairing
does not share the localization properties of the BdG wave functions
and the quasiperiodicity of the underlying AAH system has a smaller
influence on the localization of the superconducting pairing.

\subsection{Critical temperature and zero-temperature pairing\label{subsec:Critical-temperature and zero-temperature pairing}}

\begin{figure}
\begin{centering}
\includegraphics[scale=0.19]{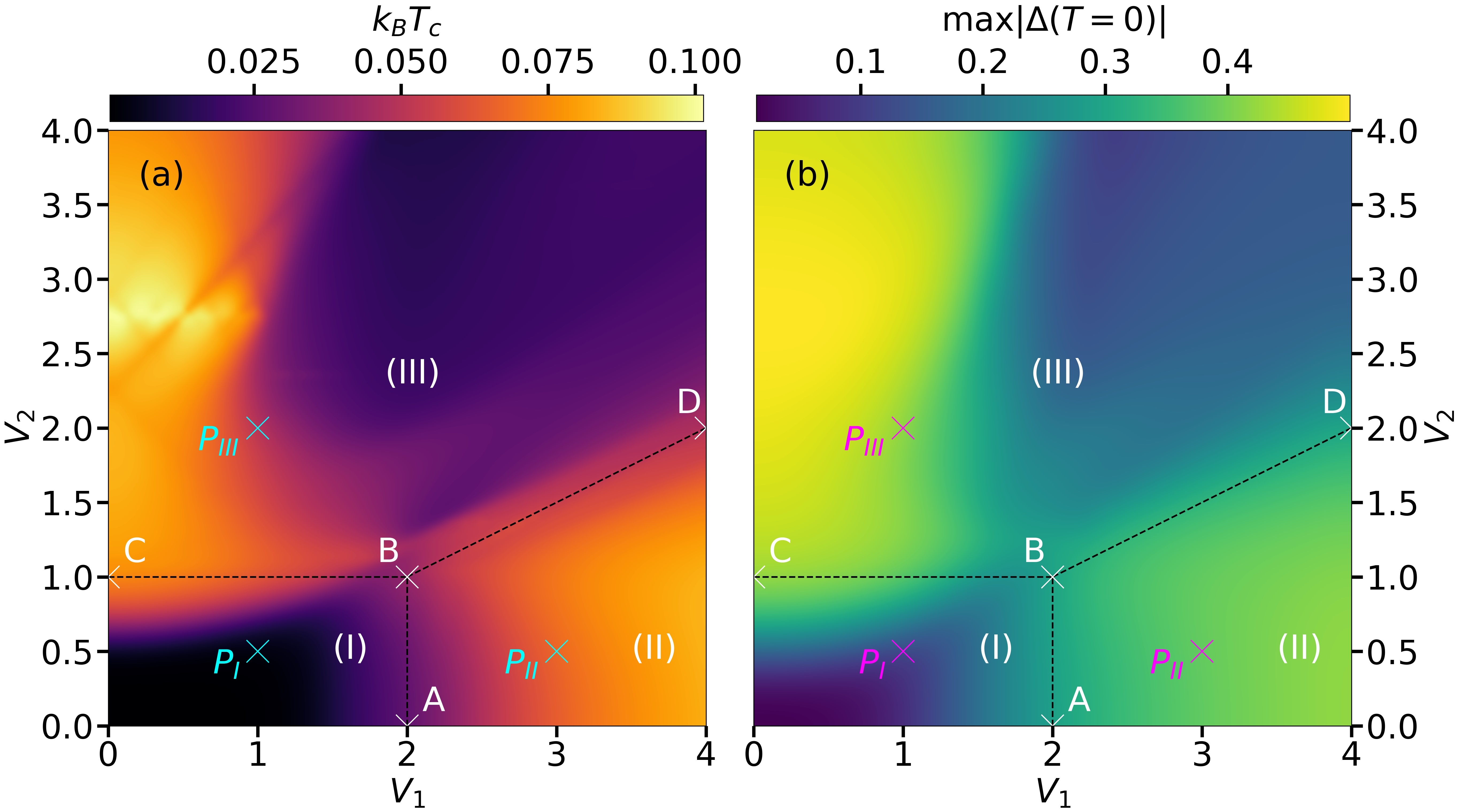}
\par\end{centering}
\begin{centering}
\includegraphics[scale=0.19]{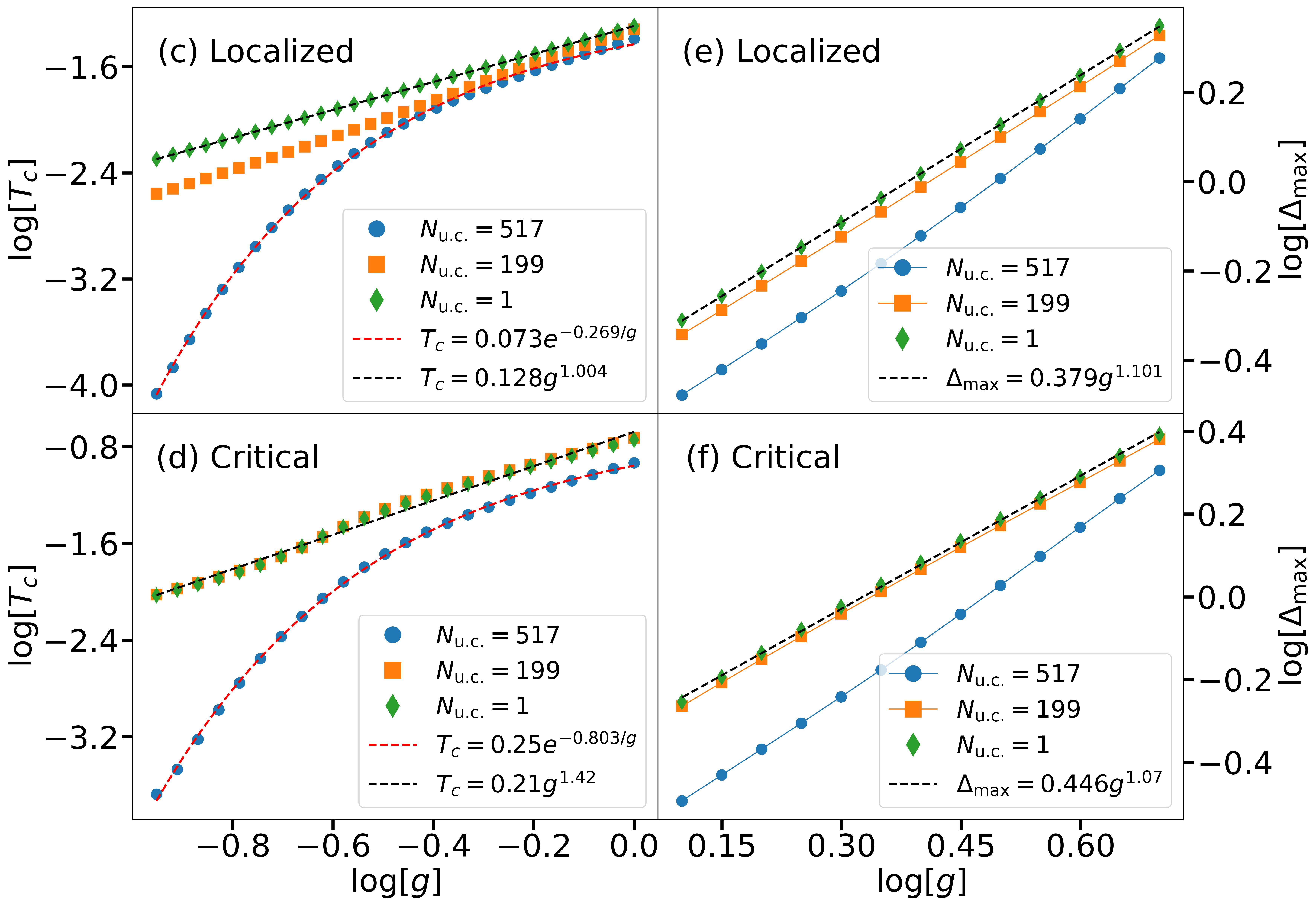}
\par\end{centering}
\centering{}\caption{(a) $\left(V_{1},V_{2}\right)$ diagram of the critical temperature,
$T_{c}$, computed with the linearized gap equation given in Eq.~\eqref{eq:Linear gap equation Anderson approximation},
for $g=0.5$ and $N=987$. (b) $\left(V_{1},V_{2}\right)$ diagram
of $\Delta_{\text{max}}$ computed with the BdG method, for $g=1$
and $N=610$. (c-f) Scaling with $g$ for different numbers of unit
cells, $N_{\text{u.c.}}\in\left\{ 1,199,517\right\} $, corresponding
to $Q_{i}\in\left\{ 1597/2584,8/13,3/5\right\} $, respectively: (c)
and (d) are the $T_{c}$ results for the localized and critical phases.
(e) and (f) are the $\Delta_{\text{max}}$ results for the localized
and critical phases. The representative points of both these phases
are marked as $P_{\text{II}}$ and $P_{\text{III}}$ in Figs.~\ref{fig:g-scaling}(a)
and \ref{fig:g-scaling}(b). The black-dashed lines correspond to
the power-law fits, while the red-dashed ones correspond to the BCS
fits.\label{fig:g-scaling}}
\end{figure}

In Fig.~\ref{fig:g-scaling}(a), we report the critical temperature
in the $\left(V_{1},V_{2}\right)$ parameter space of the generalized
AAH model, computed using the linearized gap equation given in Eq.~\eqref{eq:Linear gap equation Anderson approximation}.
For a representative value of the interaction parameter, $g=0.5$,
the critical temperature is significantly higher in the localized
phase II compared to the extended phase I. This same enhancement also
occurs in the critical phase III for small enough values of the on-site
parameter, $V_{1}\lesssim1$. However, for larger values of $V_{1}$
in the critical phase, there is a suppression of the critical temperature.
It is still higher than in the extended phase, but it attains smaller
values if compared with the localized phase. In addition to this,
we demonstrate, in Figs.~\ref{fig:g-scaling}(c) and \ref{fig:g-scaling}(d),
the power-law behavior of $T_{c}$ with respect to $g$ within the
localized and critical phases of our generalized AAH model (with different
exponents), in contrast to the typical BCS-like behavior, $T_{c}\sim\exp(-c/g)$,
(where $c$ depends on the noninteracting Hamiltonian parameters)
in the extended phase. This is in good agreement to known results
for the AAH model from Ref.~\citep{Fan2021}. We compare the incommensurate
case with results for two commensurate systems with roughly the same
number of sites ($N\simeq2584$), but with different unit cell sizes.
As shown in Figs.~\ref{fig:g-scaling}(c) and \ref{fig:g-scaling}(d),
for the commensurate system with the smallest unit cell we recover
the BCS-like scaling. When considering a commensurate system with
larger unit cell, the scaling approaches the power law behavior of
the fully incommensurate case ($N_{\text{u.c}.}=1$). We have verified
that the single-particle spectra of the commensurate and incommensurate
systems are very similar, in particular displaying comparable high
density of states near the Fermi level. Therefore, the emergence of
power law scaling cannot be attributed solely to these spectral features,
but must instead be linked to the quasiperiodic nature of the system.
Furthermore, this result also highlights the impossibility of simulating
a quasiperiodic system using commensurate approximants with small
unit cell.

We also considered the maximum of the superconducting pairing in the
real-space basis at zero temperature, $\Delta_{\text{max}}\equiv\max_{i\in\{1,..,N\}}\left|\Delta_{i}\left(T=0\right)\right|$,
as a scalar order parameter to describe the superconducting state.
In Fig.~\ref{fig:g-scaling}(b), we show a density plot of $\Delta_{\text{max}}$
in the $\left(V_{1},V_{2}\right)$ parameter space, which shows a
similar structure to the one found for the critical temperature in
Fig.~\ref{fig:g-scaling}(a). This indicates that the underlying
generalized AAH model strongly determines the behavior of these superconductivity
properties. In Figs.~\ref{fig:g-scaling}(e) and \ref{fig:g-scaling}(f)
the power-law behavior $\Delta_{\text{max}}\sim g^{\eta}$ is found
for both the the localized and critical phases. When we compare the
commensurate case with the incommensurate one, also shown in Figs.~\ref{fig:g-scaling}(e)
and \ref{fig:g-scaling}(f), we always observe the power law behavior,
contrasting with the scaling of $T_{c}$ with $g$ of Figs.~\ref{fig:g-scaling}(c)
and \ref{fig:g-scaling}(d). However, we do observe a slight enhancement
of $\Delta_{\text{max}}$ as the incommensurate case is approached.

\subsection{Drude weight and superfluid weight\label{subsec:Drude-Weight-and-Superfluid-Weight}}

\begin{figure}
\begin{centering}
\includegraphics[scale=0.19]{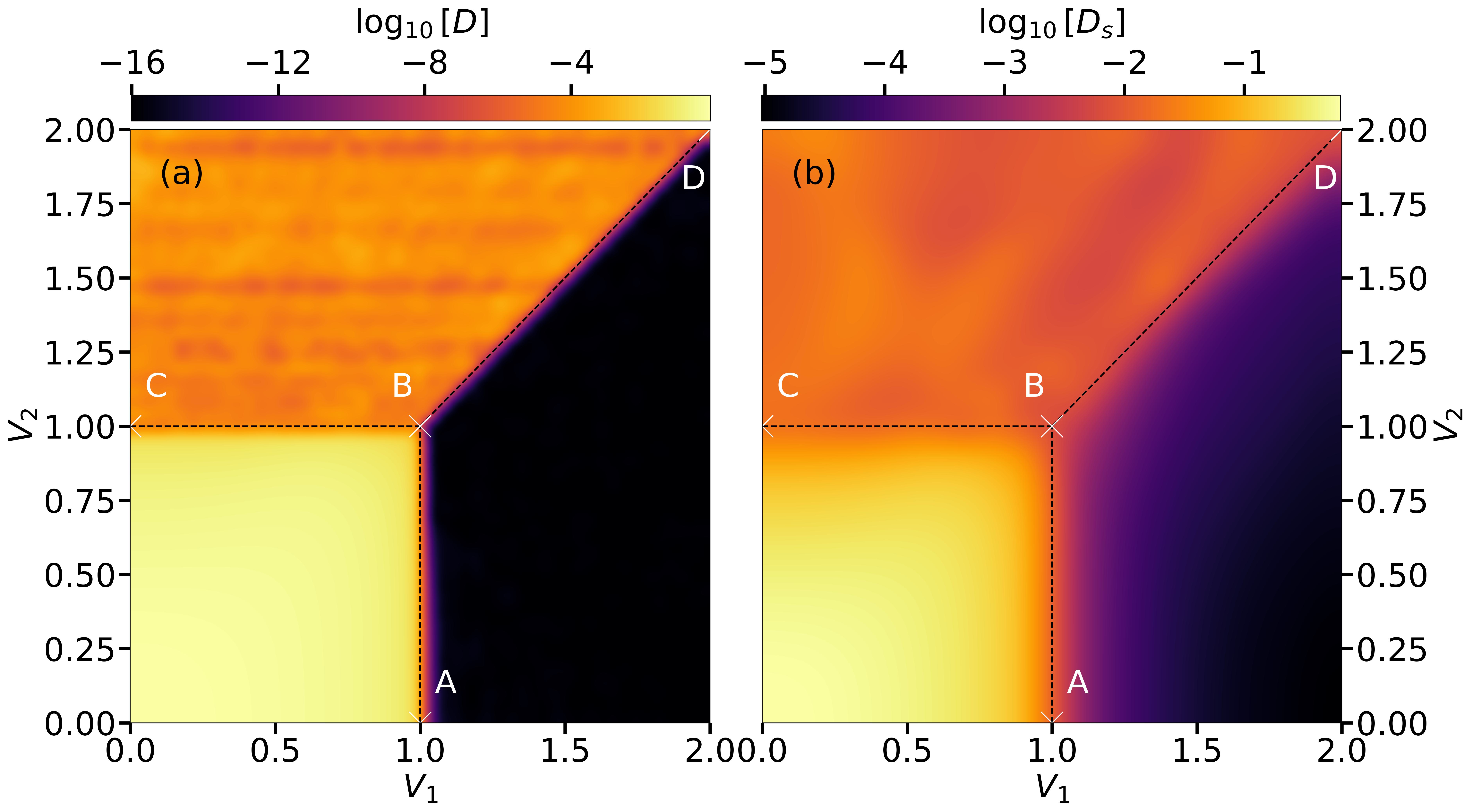}
\par\end{centering}
\begin{centering}
\includegraphics[scale=0.19]{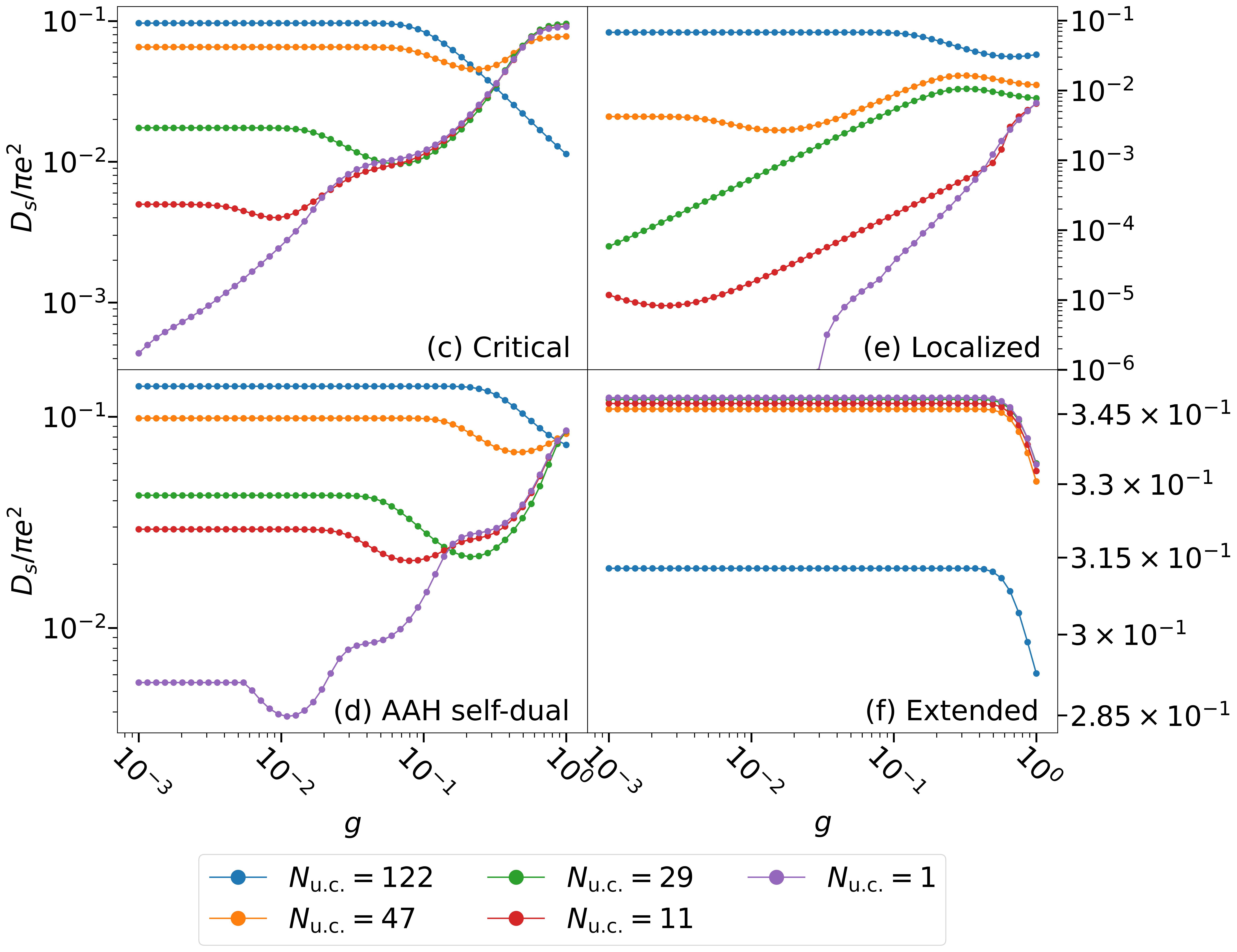}
\par\end{centering}
\centering{}\caption{(a) Drude weight, $D$, in the $\left(V_{1},V_{2}\right)$ parameter
space, computed with Eq.~\ref{eq:Superfluid weight linear response theory}
at $T=g=0$, in the incommensurate limit with system size $N=610$.
(b) Superfluid weight, $D_{s}$, in the $\left(V_{1},V_{2}\right)$
parameter space, also computed with Eq.~\ref{eq:Superfluid weight linear response theory}
at $T=0$ and finite pairing $g=0.1$, in the incommensurate limit
with system size $N=610$. (c-f) Small-coupling $g$ scaling of the
superfluid weight for different unit cell sizes $N_{\text{u.c.}}$
but system size fixed at $N\approx610$. The representative points
of both these phases are marked as $P_{\text{I}}$, $P_{\text{II}}$
and $P_{\text{III}}$ in Figs.\ref{fig:g-scaling}(a) and \ref{fig:g-scaling}(b).
(c) Critical phase at point $P_{\text{III}}$ with $\left(V_{1},V_{2}\right)=\left(1,2\right)$.
(d) AAH self-dual point with $\left(V_{1},V_{2}\right)=\left(2,0\right)$.
(e) Localized phase at point $P_{\text{II}}$ with $\left(V_{1},V_{2}\right)=\left(3,0.5\right)$.
(f) Extended phase at point $P_{\text{I}}$ with $\left(V_{1},V_{2}\right)=\left(1,0.5\right)$.
\label{fig:Ds}}
\end{figure}

In the previous sections, we showed that both the critical temperature,
computed with the linearized gap equation, and the zero-temperature
superconducting pairing, computed with the BdG method, are enhanced
inside the critical and localized phases of the $\left(V_{1},V_{2}\right)$
diagram. However, we know that in the normal state the AAH model is
an Anderson insulator in the localized phase. Therefore, it is not
clear that a finite superconducting gap is sufficient to ensure that
the system can sustain a superconducting current. It is known that
in 1D and 2D, phase fluctuations can destroy superconductivity, a
fact that we do not consider here, since we are mostly concerned in
comparing incommensurate systems with commensurate ones. In order
to determine if the AAH model actually supports a supercurrent in
the presence of superconducting pairing, at mean-field level, we have
computed the superfluid weight, $D_{s}$. The superfluid weight serves
as a measure of the phase rigidity of the system. The persistent flow
in our 1D system can be induced by constructing a ring system, i.e.,
connecting both ends of the generalized AAH chain with periodic boundary
conditions, and threading a flux through it. The superfluid weight
will then describe the current response to the static vector potential
along the ring, associated with this flux. Scalapino \textit{et}\textit{
}\textit{al}\textit{.} \citep{Scalapino1993} argued that a criteria
for the distinction between metal, insulator, and superconductor can
be constructed by studying the superfluid weight $D_{s}$ and the
Drude weight $D$, which is the current response to a uniform electric
field in the limit of zero frequency. Notice that while $D_{s}$ describes
a thermodynamic static response, $D$ describes a dynamical response.
In linear response theory, both these quantities can be computed in
1D in terms of the lattice average of the $xx$ component of the diamagnetic
tensor $\left\langle K_{xx}\right\rangle $ and appropriate zero-frequency
and infinite-wavelength limits of the current-current correlation
function. At $T=0$ and for finite coupling $g$, the Drude weight
equals the superfluid weight \citep{Giamarchi1995,Kirchner1999,Shastry2006,Mukerjee2008,Kiely2022}
and will be computed with the linear response theory result \citep{Scalapino1993},
\begin{equation}
\frac{D_{s}}{\pi e^{2}}=-\left\langle K_{xx}\right\rangle +\Lambda_{xx}(i\omega_{n}=0),\label{eq:Superfluid weight linear response theory}
\end{equation}
where $\Lambda_{xx}(i\omega_{n}=0)$ is the $xx$ component of the
Matsubara current-current correlation function at zero-frequency.
We will talk about a superconductor when the superfluid/Drude weight
is finite and an insulator when it is zero. An explicit expression
for $D_{s}$ in terms of the BdG eigenstates is provided in Appendix~1.
At $T=0$ and $g=0$, it can be shown that we always have $D_{s}=0$
for the noninteracting generalized AAH model, so it only makes sense
to talk about the Drude weight. A discussion on the differences between
$D$ and $D_{s}$ is given in Appendix~2. It is shown that $D$ and
$D_{s}$ differ only by a Fermi surface term which, at $T=g=0$, leads
to $D_{s}=0$. We show in Appendix~3 that, for $g=0$, $D$ is always
finite in the extended phase, so that we have a normal metal, but
converges to zero as we increase system size in the critical and localized
phases, characterizing an insulator ground-state. We can see in Fig.~\ref{fig:Ds}(a)
the obtained $D$ in the $\left(V_{1},V_{2}\right)$ parameter space,
where we have a vanishing Drude weight in the critical and localized
regions. Even though $D$ vanishes exponentially with system size
in the localized phase, it vanishes at a much slower rate in the critical
phase, which results in a small, yet finite, value for the system
size chosen in Fig.~\ref{fig:Ds}(a).

\begin{figure}[H]
\begin{centering}
\includegraphics[scale=0.19]{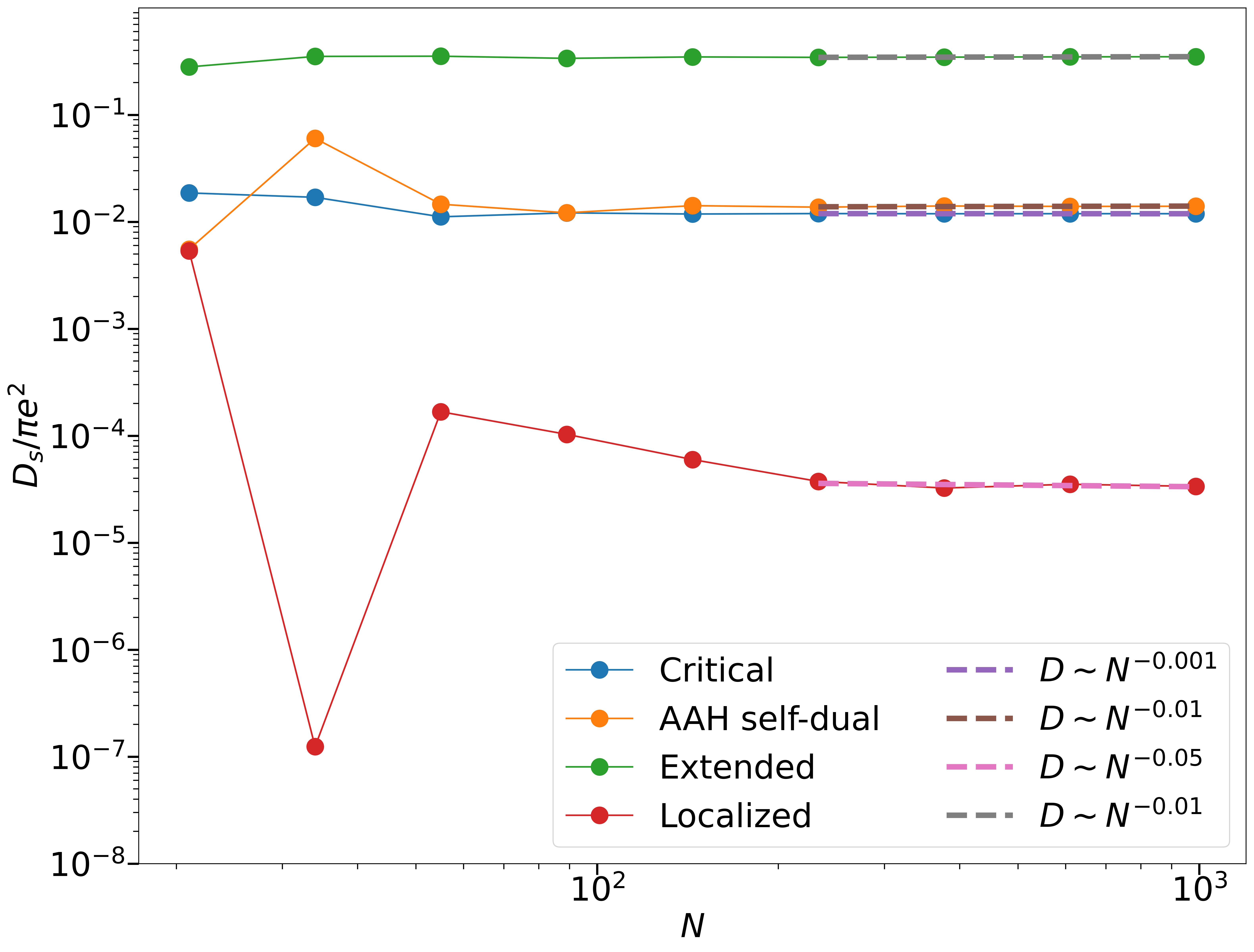}
\par\end{centering}
\caption{(a) Scaling with system size $N$ of the superfluid weight, $D_{s}$,
computed with linear response theory for $g=0.1$. We plot curves
in the three phases of the generalized AAH model, points $P_{\text{I}}$,
$P_{\text{II}}$ and $P_{\text{III}}$, and for the AAH self-dual
point, $\left(V_{1},V_{2}\right)=\left(2,0\right)$. \label{fig:Ds scaling quasiperiodic systems}}
\end{figure}

Surprisingly, we observe that, for any finite $g$, the superfluid
weight $D_{s}$ converges to a finite value in the localized and critical
phases, both in the quasiperiodic and periodic case. We show in Fig.~\ref{fig:Ds scaling quasiperiodic systems}
the finite-size analysis of $D_{s}$ in all these phases, where it
is clear that $D_{s}$ saturates to a finite value in the thermodynamic
limit at the level of our mean-field analysis. This allows us to draw
the conclusion that the system exhibits finite phase coherence in
all phases for finite $g$. Although finite, the phase coherence is
greatly suppressed in the critical phase and even more so in the localized
phase and, as we see in Fig.~\ref{fig:Ds}(b), for small values of
$g$, $D_{s}$ shares a similar $\left(V_{1},V_{2}\right)$ diagram
with $D$.

Since our focus lies in identifying signatures of quasiperiodicity
in the superconductivity properties of our system, we now turn to
the study of the superfluid weight as a function of $g$ for different
numbers of unit cells $N_{\text{u.c.}}$, just as we did for the critical
temperature and the zero-temperature pairing. We show in Figs.~\ref{fig:Ds}(c)-\ref{fig:Ds}(f)
the aforementioned scaling of $D_{s}$ with $g$ for approximants
with different $N_{\text{u.c.}}$ with the total system size fixed
at $N\approx610$. The system with $N_{\text{u.c.}}=122$ unit cells
can be understood as the commensurate limit, while the system with
$N_{\text{u.c.}}=1$ unit cells, i.e., the system consisting of a
single large unit cell, corresponds to the incommensurate limit. For
intermediate and large coupling $g\apprge t$, all $D_{s}(g)$ curves
for different $N_{\text{u.c.}}$ merge into a single curve, indicating
that quasiperiodicity is irrelevant for the static response when we
leave the weak-coupling regime. Nonetheless, as we lower $g$ and
get into the weak-coupling limit, the quasiperiodic nature of the
generalized AAH model begins to manifest itself, and we can observe
a splitting of the $D_{s}(g)$ curves. It can be seen that $D_{s}$,
for a fixed low value of $g$, decreases with decreasing $N_{\text{u.c.}}$,
attaining its lowest value in the incommensurate limit. However, if
$g>0$ then it is always finite in the limit of large $N$. This is
a strong signature of the effect of quasiperiodicity, that is only
relevant in the weak-coupling limit where its effects are not masked
by the superconducting pairing. It is also interesting to notice that
in the extended phase, quasiperiodicity actually results in a small
enhancement of $D_{s}$, as seen in Fig.~\ref{fig:Ds}(f).

\section{Discussion\label{sec:Discussion}}

By studying a generalized AAH model with spin-singlet pairing, we
analyzed the interplay between superconductivity and quasiperiodicity
across extended, localized and critical quantum phases. A central
focus of our work was the direct comparison between quasiperiodic
and similar periodic approximant systems. This allowed us to demonstrate
that, in the critical phase, the critical temperature exhibits a power-law
scaling with the interaction strength $g$. This scaling reverts to
the standard BCS behavior in the periodic limit, establishing that
it is the critical nature of the quasiperiodic single-particle eigenstates,
rather than spectral features such as an enhanced density of states
alone, that drives the enhancement of superconductivity in such systems.
We further performed a comprehensive study of the superfluid weight
across the full localization phase diagram, including a finite-size
scaling analysis to determine its behavior in the thermodynamic limit.
Our results reveal that the superfluid weight remains finite in the
extended, critical, and even in the localized phases, although its
magnitude is significantly suppressed. Notably, the effects of quasiperiodicity
on the superfluid weight are most pronounced in the weak-coupling
regime, further highlighting the crucial role of the underlying single-particle
eigenstates in shaping superconducting phases.

Our findings demonstrate that quasiperiodicity can play a crucial
role in shaping the nature of superconducting states, particularly
in systems where the underlying single-particle states have nontrivial
real-space structure. Higher dimensional systems also exhibit similar
quasiperiodic-driven phases \citep{Devakul2017,Fu2020,Goncalves_2022_2DMat}.
We therefore anticipate that in these systems, quasiperiodicity effects
will also play a critical role in understanding superconducting phases,
in strong quasiperiodic regimes, such as those recently observed in
Ref.~\citep{Uri2023} for twisted trilayer graphene. We leave the
exploration of the interplay between superconductivity and quasiperiodicity
in 2D for future work.
\begin{acknowledgments}
R.O., B.A., and E.V.C. acknowledge support from Fundação para a Ciência
e a Tecnologia through Grant No. UID/04650/2025, Grant No. UID/PRR/04650/2025~\citep{UID/PRR/04650/2025},
Grant No. UID/PRR2/04650/2025~\citep{UID/PRR2/04650/2025}, and project
QPIMoireS~\citep{2023.17745.ICDT}.

P.R. acknowledges support by Fundação para a Ciência e a Tecnologia
through Grant No. UID/04540/2025~\citep{UID/PRR/04540/2025} to the
I\&D unit Centro de Física e Engenharia de Materiais Avançados (CeFEMA)
and project SCALE-QLT~\citep{2024.16192.PEX}. M.G. is supported
by a postdoctoral research fellowship at the Princeton Center for
Theoretical Science.

We finally acknowledge the Tianhe-2JK cluster at the Beijing Computational
Science Research Center, the Bob\ensuremath{\mid}Macc supercomputer
(Computational Project CPCA/A1/470243/2021) and the OBLIVION supercomputer
(Projects HPCUE/A1/468700/2021, 2022.15834.CPCA.A1 and 2022.15910.CPCA.A1).
The OBLIVION supercomputer is at the High Performance Computing Center,
University of Évora, and is funded by the ENGAGE SKA Research Infrastructure
(Reference POCI-01-0145-FEDER-022217 - COMPETE 2020), the Fundação
para a Ciência e a Tecnologia, the BigData@UE project (Reference ALT20-03-0246-FEDER-000033
- FEDER) and the Alentejo 2020 Regional Operational Program. Computer
assistance was provided by the Computational Science Research Center,
Bob\ensuremath{\mid}Macc and OBLIVION support teams.
\end{acknowledgments}

\section*{Appendix\label{sec:Appendix}}

\subsection*{1. $D_{s}$ in terms of the BdG solutions\label{sec:Appendix 1}}

As we mentioned in Sec.~\ref{subsec:Drude-Weight-and-Superfluid-Weight},
we compute the superfluid weight with the linear response theory result
(see Ref.~\citep{Zhu2016} for a derivation) 
\begin{equation}
\frac{D_{s}}{\pi e^{2}}=-\left\langle K_{xx}\right\rangle +\Lambda_{xx}(q\rightarrow0,i\omega_{n}=0),\label{eq:Ds LRT}
\end{equation}
where the $xx$-component of the diamagnetic tensor, $K_{xx}$, is
the second derivative of the Hamiltonian with respect to the vector
potential. For a 1D nearest-neighbor tight-binding Hamiltonian minimally
coupled to a vector potential via Peierls substitution, $\left\langle K_{xx}\right\rangle $
can be written as
\begin{align}
\left\langle K_{xx}\right\rangle  & \equiv\frac{1}{2N}\left\langle \left.\frac{\partial^{2}H}{\partial A_{x}\partial A_{x}}\right|_{A_{x}=0}\right\rangle \nonumber \\
 & =-\frac{1}{2N}\sum_{m,\sigma}t_{m}\left[\left\langle c^{\dagger}_{m,\sigma}c_{m+1,\sigma}\right\rangle +\left\langle c^{\dagger}_{m+1,\sigma}c_{m,\sigma}\right\rangle \right],\label{eq:Diamagnetic term}
\end{align}
with $t_{m}$ being the hopping term from site $m$ to site $m+1$,
where for the considered generalized AAH model takes the form $t_{m}=t+V_{2}\cos\left(2\pi Qm+\pi Q\right)$,
$N$ is the total system size, and $\left\langle ...\right\rangle $
is the average with respect to the unperturbed thermodynamic equilibrium
state. The second term in Eq.~\eqref{eq:Ds LRT} is the current-current
correlation function in Matsubara frequency space
\begin{equation}
\Lambda_{xx}\left(q,i\omega_{n}\right)=-\frac{1}{N}\intop^{\beta}_{0}d\tau e^{i\omega_{n}\tau}\left\langle J^{P}_{x}\left(q,\tau\right)J^{P}_{x}\left(-q,0\right)\right\rangle _{0},\label{eq:Current-current correlation function}
\end{equation}
 that is written in terms of the Fourier transform of the paramagnetic
current density for a 1D system
\begin{multline}
J^{P}_{x}\left(q,\tau\right)=\\
i\sum_{m,\sigma}e^{-iqm}t_{m}\left[c^{\dagger}_{m+1,\sigma}\left(\tau\right)c_{m,\sigma}\left(\tau\right)-c^{\dagger}_{m,\sigma}\left(\tau\right)c_{m+1,\sigma}\left(\tau\right)\right]\label{eq:Paramagnetic current density}
\end{multline}

After performing a canonical transformation, the superfluid weight
can be written in terms of the BdG eigenstates \eqref{eq:BdG canonical transformation}
as 
\begin{align}
\frac{D_{s}}{\pi e^{2}} & =\sum_{\mu}\left[\left\langle K^{V}_{\mu}\right\rangle f\left(-E_{\mu}\right)+\left\langle K^{U}_{\mu}\right\rangle f\left(E_{\mu}\right)\right]\nonumber \\
 & +\frac{1}{2N}\sum_{\mu,\nu}\chi\left(E_{\mu},E_{\nu}\right)J^{U,U}_{\mu,\nu}\left[\left(J^{U,U}_{\mu,\nu}\right)^{*}+J^{V,V}_{\mu,\nu}\right]\nonumber \\
 & +\frac{1}{2N}\sum_{\mu,\nu}\chi\left(-E_{\mu},-E_{\nu}\right)J^{V,V}_{\mu,\nu}\left[\left(J^{V,V}_{\mu,\nu}\right)^{*}+J^{U,U}_{\mu,\nu}\right]\nonumber \\
 & +\frac{1}{2N}\sum_{\mu,\nu}\chi\left(-E_{\mu},E_{\nu}\right)J^{V,U}_{\mu,\nu}\left[\left(J^{V,U}_{\mu,\nu}\right)^{*}+J^{U,V}_{\mu,\nu}\right]\nonumber \\
 & +\frac{1}{2N}\sum_{\mu,\nu}\chi\left(E_{\mu},-E_{\nu}\right)J^{U,V}_{\mu,\nu}\left[\left(J^{U,V}_{\mu,\nu}\right)^{*}+J^{V,U}_{\mu,\nu}\right].\label{eq:BdG form of superfluid weight}
\end{align}
where the eigenvalue-dependent kernel is defined as
\begin{equation}
\chi\left(E_{\mu},E_{\nu}\right)\equiv\frac{f\left(E_{\mu}\right)-f\left(E_{\nu}\right)}{E_{\mu}-E_{\nu}},\label{eq:Eigenvalue kernel}
\end{equation}
and the electron/hole matrix elements of the diamagnetic tensor are
written as

\begin{align}
K^{U}_{\mu} & \equiv\left\langle U_{\mu}\right|K_{xx}\left|U_{\mu}\right\rangle \nonumber \\
 & =\sum_{m,\sigma}t_{m}\left(U^{*}_{m,\sigma;\mu}U_{m+1,\sigma;\mu}+U^{*}_{m+1,\sigma;\mu}U_{m,\sigma;\mu}\right),\label{eq:Electron diamagnetic matrix elements}
\end{align}

\begin{align}
K^{V}_{\mu} & \equiv\left\langle V^{*}_{\mu}\right|K_{xx}\left|V^{*}_{\mu}\right\rangle \nonumber \\
 & =\sum_{m,\sigma}t_{m}\left(V_{m,\sigma;\mu}V^{*}_{m+1,\sigma;\mu}+V_{m+1,\sigma;\mu}V^{*}_{m,\sigma;\mu}\right),\label{eq:Hole diamagnetic matrix elements}
\end{align}
while the paramagnetic current density matrix elements are given by
\begin{align}
J^{U,U}_{\mu,\nu} & \equiv\left\langle U_{\mu}\right|J^{P}_{x}\left(q=0\right)\left|U_{\mu}\right\rangle \nonumber \\
 & =\sum_{m,\sigma}t_{m}\left(U^{*}_{m+1,\sigma;\mu}U_{m,\sigma;\nu}-U^{*}_{m,\sigma;\mu}U_{m+1,\sigma;\nu}\right),\label{eq:A-matrix}\\
J^{V,V}_{\mu,\nu} & \equiv\left\langle V^{*}_{\mu}\right|J^{P}_{x}\left(q=0\right)\left|V^{*}_{\mu}\right\rangle \nonumber \\
 & =\sum_{m,\sigma}t_{m}\left(V_{m+1,\sigma;\mu}V^{*}_{m,\sigma;\nu}-V_{m,\sigma;\mu}V^{*}_{m+1,\sigma;\nu}\right),\label{eq:B-matrix}\\
J^{V,U}_{\mu,\nu} & \equiv\left\langle V^{*}_{\mu}\right|J^{P}_{x}\left(q=0\right)\left|U_{\mu}\right\rangle \nonumber \\
 & =\sum_{m,\sigma}t_{m}\left(V_{m+1,\sigma;\mu}U_{m,\sigma;\nu}-V_{m,\sigma;\mu}U_{m+1,\sigma;\nu}\right),\label{eq:C-matrix}\\
J^{U,V}_{\mu,\nu} & \equiv\left\langle U_{\mu}\right|J^{P}_{x}\left(q=0\right)\left|V^{*}_{\mu}\right\rangle \nonumber \\
 & =\sum_{m,\sigma}t_{m}\left(U^{*}_{m+1,\sigma;\mu}V^{*}_{m,\sigma;\nu}-U^{*}_{m,\sigma;\mu}V^{*}_{m+1,\sigma;\nu}\right).\label{eq:D-matrix}
\end{align}

\subsection*{2. Difference between $D$ and $D_{s}$\label{sec:Appendix 2}}

The goal of this section is to clarify the difference between the
Drude weight, $D$, and the superfluid weight, $D_{s}$. In particular,
we will show that they differ by a Fermi surface term associated with
the manifold of degenerate states. It is a well known result in the
literature \citep{Scalapino1993,Kirchner1999,Giamarchi1995} that
both $D$ and $D_{s}$ can be written in terms of the noncommuting
limits $q\rightarrow0$ and $i\omega_{n}\rightarrow0$ of the current-current
correlation function $\Lambda(q,i\omega_{n})$. An alternative way
to think about this issue, is to essentially look at $D_{s}$ as a
thermodynamic response of the system which can be written as we did
in Eq.~\eqref{eq:Ds LRT} in terms of the current-current correlation
function in Matsubara frequency space, while $D$ can be thought as
a dynamical response and can be written in terms of the retarded current-current
correlation $\Lambda^{\text{ret}}_{xx}\left(q,\omega\right)$ as 
\begin{equation}
\frac{D}{\pi e^{2}}=-\left\langle K_{xx}\right\rangle +\Lambda^{\text{ret}}_{xx}\left(q=0,\omega\rightarrow0\right),\label{eq:D LRT}
\end{equation}
where

\begin{multline}
\Lambda^{\text{ret}}_{xx}\left(q,\omega\right)\equiv\\
\frac{-i}{N}\int^{+\infty}_{0}dte^{i\left(\omega+i0^{+}\right)t}\left\langle \left[J^{P}_{x}\left(q,t\right),J^{P}_{x}\left(-q,0\right)\right]\right\rangle \label{eq:Retarded current-current correlation function}
\end{multline}

Instead of doing the following calculations in Nambu space, we will
be pragmatic and work with a generic quadratic Hamiltonian with no
anomalous terms. In this way, the notation will be simpler and the
results are easily generalized to expressions like Eq.~\eqref{eq:BdG form of superfluid weight},
since the difference between $D$ and $D_{s}$, as we will show, lies
in the inclusion of the degenerate states in the sums over eigenstates.
In the eigenbasis $\left\{ a_{\mu}\right\} $ of the Hamiltonian,
the 1D paramagnetic current density in Eq.~\eqref{eq:Paramagnetic current density}
is given by 
\begin{equation}
J^{P}_{x}\left(q,\tau\right)=\sum_{\mu,\nu}J_{\mu,\nu}\left(q\right)a^{\dagger}_{\mu}\left(\tau\right)a_{\nu}\left(\tau\right),\label{eq:1D paramagnetic current eigenbasis}
\end{equation}
whose matrix elements can be written in terms of the unitary matrix
$U$ that diagonalizes the Hamiltonian, $c_{m,\sigma}=\sum_{\mu}U_{m,\sigma;\mu}a_{\mu}$,
as
\begin{multline}
J_{\mu,\nu}\left(q\right)=\\
i\sum_{m,\sigma}e^{-iqm}t_{m}\left[U^{*}_{m+1,\sigma;\mu}U_{m,\sigma;\nu}-U^{*}_{m,\sigma;\mu}U_{m+1,\sigma;\nu}\right],\label{eq:Paramagnetic matrix elements}
\end{multline}
satisfying the following property $J_{\mu,\nu}\left(-q\right)=J^{*}_{\mu,\nu}\left(q\right).$

Hence, the current-current correlation function can be written in
terms of the matrix elements of the paramagnetic current as 
\begin{multline}
\Lambda_{xx}\left(q,i\omega_{n}\right)=-\frac{1}{N}\int^{\beta}_{0}d\tau e^{i\omega_{n}\tau}\sum_{\mu,\nu}\sum_{\alpha,\beta}\\
J_{\mu,\nu}\left(q\right)J_{\alpha,\beta}\left(-q\right)e^{\left(E_{\mu}-E_{\nu}\right)\tau}\left\langle a^{\dagger}_{\mu}a_{\nu}a^{\dagger}_{\alpha}a_{\beta}\right\rangle _{0},\label{eq:Current-current Step 1}
\end{multline}
where $\left\{ E_{\mu}\right\} $ are the eigenvalues of the 1D Hamiltonian
that enter the expression after evolving the bare operators in imaginary
time. After using Wick's theorem and $\left\langle J^{P}_{x}\right\rangle _{0}=0$,
since we have no current in the system in the absence of a perturbation,
we obtain that the current-current correlation function at zero Matsubara
frequency is given by
\begin{multline}
\Lambda_{xx}\left(q,i\omega_{n}=0\right)=\\
-\frac{1}{N}\sum_{\mu,\nu}\left|J_{\mu,\nu}\left(q\right)\right|^{2}f\left(E_{\mu}\right)f\left(-E_{\nu}\right)\int^{\beta}_{0}d\tau e^{\left(E_{\mu}-E_{\nu}\right)\tau}.\label{eq:Current-current step 1}
\end{multline}
Performing the integration in imaginary time, we obtain
\begin{equation}
\int^{\beta}_{0}d\tau e^{\left(E_{\mu}-E_{\nu}\right)\tau}=\begin{cases}
\beta & \text{if }E_{\mu}=E_{\nu}\\
\frac{e^{\beta\left(E_{\mu}-E_{\nu}\right)}-1}{E_{\mu}-E_{\nu}} & \text{if }E_{\mu}\neq E_{\nu}
\end{cases},\label{eq:Property 1}
\end{equation}
Noticing that
\begin{equation}
\beta f\left(E_{\mu}\right)f\left(-E_{\mu}\right)=-f^{'}\left(E_{\mu}\right),\label{eq:Property 2}
\end{equation}
\begin{equation}
f\left(E_{\mu}\right)f\left(-E_{\nu}\right)\left[e^{\beta\left(E_{\mu}-E_{\nu}\right)}-1\right]=f\left(E_{\nu}\right)-f\left(E_{\mu}\right),\label{eq:Property 3}
\end{equation}
we can write the superfluid weight in Eq.~\eqref{eq:Ds LRT} as
\begin{align}
\frac{D_{s}}{\pi e^{2}} & =-\left\langle K_{xx}\right\rangle +\frac{1}{N}\sum_{E_{\mu}=E_{\nu}}f^{\prime}\left(E_{\mu}\right)\left|J_{\mu,\nu}\left(q\rightarrow0\right)\right|^{2}\nonumber \\
 & +\frac{1}{N}\sum_{E_{\nu}\neq E_{\mu}}\frac{f\left(E_{\mu}\right)-f\left(E_{\nu}\right)}{E_{\mu}-E_{\nu}}\left|J_{\mu,\nu}\left(q\rightarrow0\right)\right|^{2}.\label{eq:Final form of Ds for appendix B}
\end{align}

Finally, we will compute the Drude weight with the retarded current-current
correlation function given in Eq.~\eqref{eq:Retarded current-current correlation function}.
Once again in the eigenbasis of the Hamiltonian, after evolving the
operators in real time, we obtain
\begin{multline}
\Lambda^{\text{ret}}_{xx}\left(q,\omega\right)=\frac{1}{iN}\intop^{\infty}_{0}dte^{i\left(\omega+i\eta\right)t}\sum_{\mu,\nu}\sum_{\alpha,\beta}\\
J_{\mu,\nu}\left(q\right)J_{\alpha,\beta}\left(-q\right)e^{i\left(E_{\mu}-E_{\nu}\right)t}\left\langle \left[a^{\dagger}_{\mu}a_{\nu},a^{\dagger}_{\alpha}a_{\beta}\right]\right\rangle _{0},\label{eq:Retarded current-current step 1}
\end{multline}
and the expectation value of the commutator is easy to evaluate and
gives $\left[f\left(E_{\mu}\right)-f\left(E_{\nu}\right)\right]\delta_{\mu,\beta}\delta_{\nu,\alpha}$.
Hence, the terms in Eq.~\eqref{eq:Retarded current-current step 1}
with $E_{\mu}=E_{\nu}$ are automatically excluded in this formalism,
because they contribute nothing to the sum over eigenstates. We finally
obtain the expression for the Drude weight
\begin{align}
\frac{D}{\pi e^{2}} & =-\left\langle K_{xx}\right\rangle \nonumber \\
 & +\frac{1}{N}\sum_{E_{\mu}\neq E_{\nu}}\frac{f\left(E_{\mu}\right)-f\left(E_{\nu}\right)}{E_{\mu}-E_{\nu}}\left|J_{\mu,\nu}\left(q=0\right)\right|^{2}.\label{eq:Final expression for the Drude weight}
\end{align}

Equations~\eqref{eq:Final form of Ds for appendix B} for $D$ and
Eq.~\eqref{eq:Final expression for the Drude weight} for $D_{s}$
show that 
\begin{equation}
\text{\ensuremath{\frac{D_{s}}{\pi e^{2}}}}-\frac{D}{\pi e^{2}}=\frac{1}{N}\sum_{E_{\mu}=E_{\nu}}f^{\prime}\left(E_{\mu}\right)\left|J_{\mu,\nu}\left(0\right)\right|^{2},\label{eq:Difference between D and Ds}
\end{equation}
which, as advertised, imply that the difference in these two quantities
lies in a Fermi surface term, in the sense that, due to the sharp
nature of $f^{\prime}\left(E_{\mu}\right)$ for $E_{\mu}=\epsilon_{\mu}-\mu$
close to $0$, only the diagonal current matrix elements $J_{\mu,\mu}\left(0\right)$
associated with energies $\epsilon_{\mu}$ close to the chemical potential
contribute. It is interesting to discuss some limits of these definitions.
From Eq.~\eqref{eq:Difference between D and Ds}, it is clear that
for a gaped system (either for an insulator or a superconductor) at
$T=0$, the Fermi surface term vanishes and we obtain $D_{s}=D$,
which is consistent with the known result from Scalapino \textit{et
al.} \citep{Scalapino1993}. However, at $T=0$ and $g=0$ we will
find $D_{s}=0$ due to a finite contribution from diagonal current
matrix elements $J_{\mu,\mu}\left(0\right)$ for $E_{\mu}=0$. For
$T>T_{c}$, the system is no longer a superconductor and we will have
$D_{s}=0$. Whether $D=0$ or $D\neq0$ at these temperatures depends
if the system is a metal or an insulator.

\subsection*{3. Finite-size analysis of $D$ and $D_{s}$\label{sec:Appendix 3}}

At $T=0$, we always have that $D_{s}=D$, provided that our system
is gaped. This always the case for spin-singlet superconductivity
provided that $g\neq0$. However, for $g=0$ our system cannot, of
course, display superconductivity at the mean-field level, i.e., it
does not have a finite gap $\Delta$ and, therefore, we always have
$D_{s}=0$. Thus, when we compare the finite-size analysis of systems
with $g\neq0$ and $g=0$ we will talk about the superfluid weight,
$D_{s}$, and the Drude weight, $D$, respectively.

\begin{figure}[H]
\begin{centering}
\includegraphics[scale=0.19]{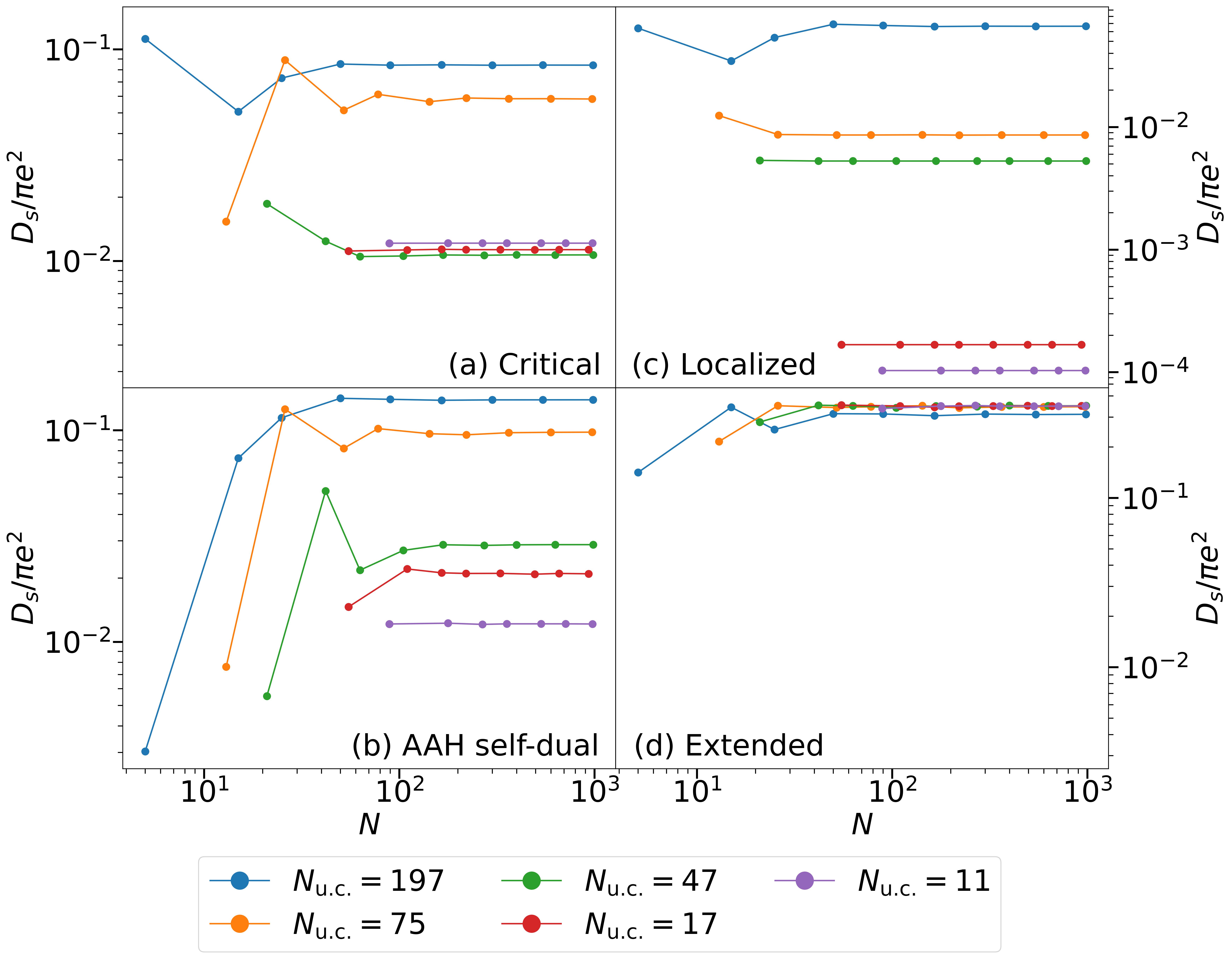}
\par\end{centering}
\centering{}

\caption{Scaling of $D_{s}$ with total system size up to $N\approx610$, for
different periodic systems with different numbers of unit cells $N_{\text{u.c.}}$.
(a) Critical phase at point $P_{\text{III}}$ with $\left(V_{1},V_{2}\right)=\left(1,2\right)$.
(b) AAH dual point with $\left(V_{1},V_{2}\right)=\left(2,0\right)$.
(c) Localized phase at point $P_{\text{II}}$ with $\left(V_{1},V_{2}\right)=\left(3,0.5\right)$.
(d) Extended phase at point $P_{\text{I}}$ with $\left(V_{1},V_{2}\right)=\left(1,0.5\right)$\label{fig:Ds scaling periodic systems}}
\end{figure}

We show in Fig.~\ref{fig:Ds scaling periodic systems} the scaling
of the superfluid weight $D_{s}$ with total system size $N$. For
periodic systems, we find that $D_{s}$ converges to a finite value
when we increase $N$ in all phases of the generalized AAH model:
extended (I), localized (II) and critical (III). This is to be expected,
since for low values of $N_{\text{u.c.}}$ the eigenstates of the
system still have an extended nature in all phases. Given that we
are far from the incommensurate limit, it is natural that $D_{s}$
converges to a finite value, reflecting the fact that the system is
not an insulator for periodic modulations of the potential or the
hoppings.

\begin{figure}[H]
\begin{centering}
\includegraphics[scale=0.19]{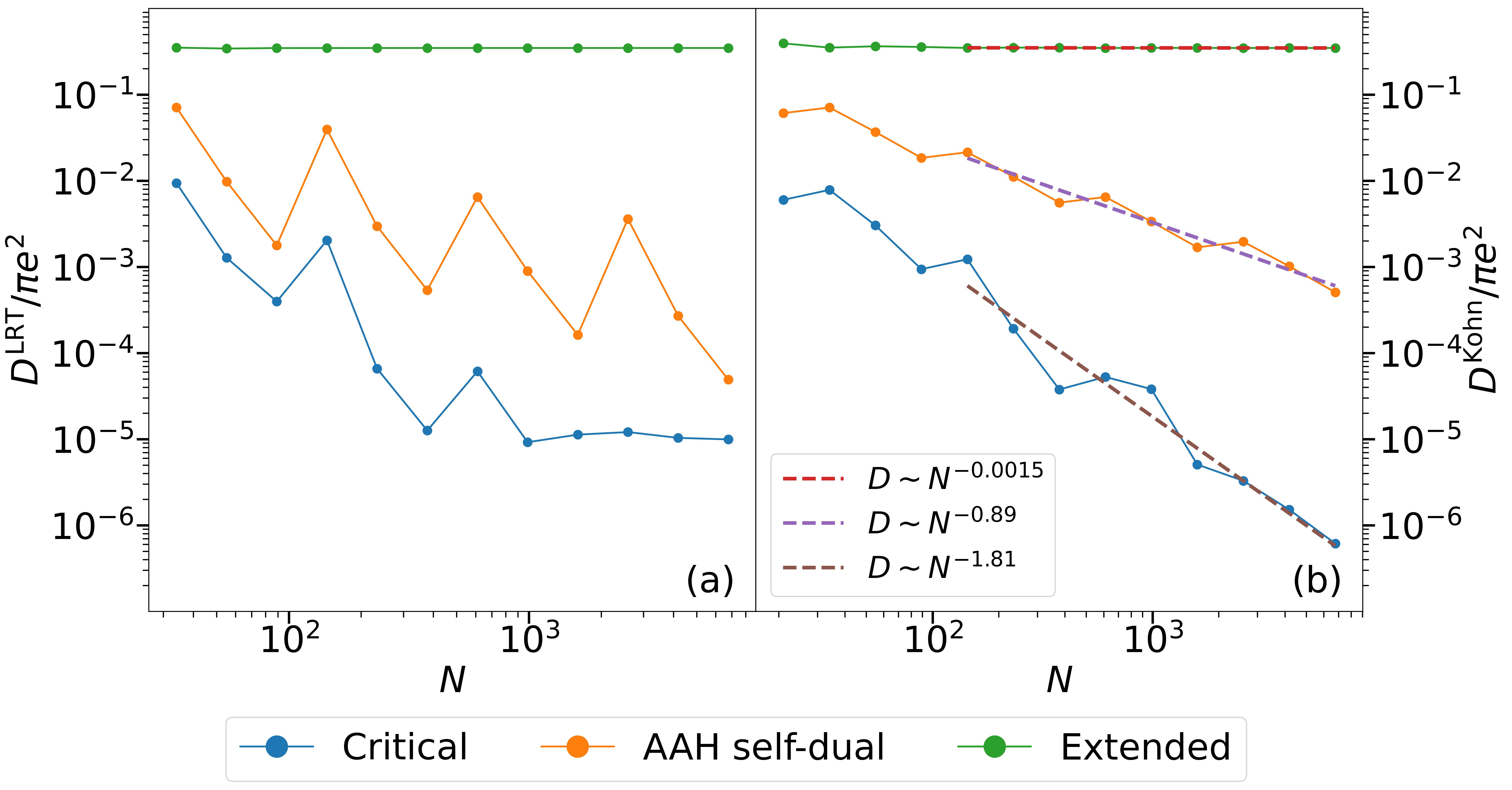}
\par\end{centering}
\caption{(a) Scaling with system size $N$ of the Drude weight computed with
linear response theory, $D^{\text{LRT}}$, for $g=0$. (b) Scaling
of the Drude weight computed with Kohn's method, $D^{\text{Kohn}}$,
for $g=0$. We plot curves in two phases of the generalized AAH model,
points $P_{\text{I}}$ and $P_{\text{III}}$, and for the AAH self-dual
point, $\left(V_{1},V_{2}\right)=\left(2,0\right)$. We do not plot
the curve for the localized phase (II) for better reading of the figure,
since $D$ scales exponentially to zero and reaches machine precision
for small system sizes. \label{fig:D scaling quasiperiodic systems}}
\end{figure}

However, we obtain an interesting result when we compare the $g=0$
and the $g\neq0$ cases for quasiperiodic systems. We do a finite-size
analysis of quasiperiodic systems by setting $N_{\text{u.c.}}=1$
and studying the scaling of $D_{s}$, as shown in Fig.~\ref{fig:Ds scaling quasiperiodic systems},
or $D$ with increasing $N$, as can be seen in Fig.~\ref{fig:D scaling quasiperiodic systems}.
For $g=0$, we find that $D$ saturates to a finite value in the extended
phase (I), goes exponentially to zero in the localized phase (II),
and also goes to zero in the critical phase (III) but with a much
slower convergence, as shown in Fig.~\ref{fig:D scaling quasiperiodic systems}(a).
This is to be expected, since we know the noninteracting quasiperiodic
system to be an insulator in phases (II) and (III) but a metal in
phase (I). The nontrivial behavior occurs when we look at the scaling
of $D_{s}$ at finite $g$, presented in Fig.~\ref{fig:Ds scaling quasiperiodic systems}.
In this case, we find that $D_{s}$ converges to a finite value in
all phases of the generalized AAH model. Hence, we no longer obtain
an insulating behavior in phases (II) and (III) for our quasiperiodic
system, provided that we have a finite interaction parameter $g$.

A note should be made that the linear response theory result for the
convergence of $D$ with system size for the critical phase (III),
as seen in Fig.~\ref{fig:D scaling quasiperiodic systems}(a), is
not clear about whether $D$ saturates to a finite value or goes to
$0$. To get a more assertive answer to this question, we also did
a finite-size analysis of $D$ computed with Kohn's method \citep{Kohn1964}.
In this method, the Drude weight is computed as the second derivative
of the ground-state energy with respect to the flux or twist in the
boundary conditions,
\begin{equation}
D=\left.N\frac{\partial^{2}E_{0}}{\partial\phi^{2}}\right|_{\phi=\phi_{0}},\label{eq:Drude weight Kohn's method}
\end{equation}
where $E_{0}$ is the ground-state energy and $\phi_{0}$ is the twist
for which the dispersion $E_{0}\left(\phi\right)$ has a minimum.
With this alternative method, we were able to confirm that $D$ also
scales to zero in the critical phase (III), as seen in Fig.~\ref{fig:D scaling quasiperiodic systems}(b).

\bibliographystyle{apsrev4-2}
\bibliography{Refs}

\end{document}